\documentstyle[11pt]{article}
\renewcommand{\baselinestretch}{1.5}

\begin{document}

\begin{flushright}
BRX TH-324, IHES/P/92/36, ADP-92-180/M5
\end{flushright}

\begin{flushright}
June, 1992
\end{flushright}

\vspace{.1in}

\begin{center}
{\Large{\bf Nonsymmetric Gravity Theories: Inconsistencies and a Cure }}

\vspace{.2in}
\renewcommand{\baselinestretch}{1}
\small
\normalsize
{\it T. Damour\\
Institut des Hautes Etudes Scientifiques\\
91440 Bures sur Yvette\\
and\\
D.A.R.C., CNRS -- Observatoire de Paris\\
92195 Meudon, France}

\vspace{.2in}

{\it S. Deser \\
Physics Department\\
Brandeis University\\
Waltham, MA 02254, USA}

\vspace{.2in}

{\it J. McCarthy \\
Department of Physics and Mathematical Physics\\
University of Adelaide\\
GPO Box 498, Adelaide, SA 5001, Australia}

\end{center}
\renewcommand{\baselinestretch}{1.5}
\small
\normalsize

\vspace{.2in}

Motivated by the apparent dependence of string $\sigma$--models
on the sum of spacetime metric and antisymmetric tensor
fields, we reconsider gravity theories constructed from a
nonsymmetric metric.  We first show
that all such ``geometrical" theories homogeneous in
second derivatives violate  standard physical requirements:
ghost-freedom, absence of algebraic inconsistencies or continuity
of degree-of-freedom content.  This no-go result applies in particular
to the old unified theory of Einstein and its recent avatars.
However, we find that the addition of nonderivative, ``cosmological''
terms formally restores consistency by giving a mass to the
antisymmetric tensor field, thereby transmuting it into a fifth-force-like
massive vector but with novel possible matter couplings.
The resulting macroscopic models also exhibit
``van der Waals''-type gravitational effects, and may provide useful
phenomenological foils to general relativity.

\vspace{.1in}
\renewcommand{\theequation}{1.\arabic{equation}}
\setcounter{equation}{0}

\noindent{\bf I. ~Introduction.}

It is a remarkable historical coincidence that modern string
theory can be interpreted as reviving an ancient attempt at geometric
unification of forces.  Specifically, we note that (ignoring the
dilaton for simplicity) the string $\sigma$--model action in conformal
gauge is just
\begin{equation}
I = \int d^{2}z (G_{\mu\nu}(X) + B_{\mu\nu}(X) )
\partial_z\,X^\mu \,\partial_{\bar{z}} \, X^\nu
\label{Aa}
\end{equation}
in terms of local complex world-sheet coordinates $z,\bar{z}$.
Here $X^\mu$ is the embedding coordinate of the string in a
$D$-dimensional
spacetime endowed with the (symmetric) metric $G_{\mu\nu}$ and the
antisymmetric tensor field $B_{\mu\nu}$.  The action (\ref{Aa})
thus depends only on the ``unified'' combination
\begin{equation}
g_{\mu\nu} \equiv G_{\mu\nu} + B_{\mu\nu} \; ,
\label{Ab}
\end{equation}
and we are led to consider field theories of gravity constructed
geometrically from the non-symmetric metric $g_{\mu\nu}$.  One such
model was first discussed in 1925 by Einstein
\cite{Ein} in an attempt to unify gravity and electromagnetism
($B_{\mu\nu}$
was to be related to the Maxwell field strength).  Further treatment by
Einstein and his collaborators \cite{EnO},
Schr\"{o}dinger \cite{ES}, and others \cite{L55} \cite{T55} \cite{MTV}, led to
considerable
elaboration of its mathematical structure.  More recently, Moffat \cite{Moff}
has proposed that $B_{\mu\nu}$ be interpreted as a new field
coupled to macroscopic currents.  Following recent usage, we shall
refer to this specific theory as NGT
(``Nonsymmetric Gravity Theory''), but we shall also analyze the whole
class of geometric non-symmetric theories.

  {\it A priori}, nonsymmetric geometrization suffers from a number of
weaknesses.  Not least is the absence
of a natural way to unify the two initial building blocks
$G_{\mu\nu}$ and $B_{\mu\nu}$ into a single entity.  This
objection was originally raised by Weyl \cite{Weyl} and Pauli \cite{Pauli}
in connection with Einstein's attempts; it states that because
$g_{\mu\nu}$ is a reducible
representation of diffeomorphisms, there is no real meaning
in saying that a theory is expressed ``solely in terms of $g_{\mu\nu}$".
Indeed, any theory involving $G_{\mu\nu}$ and $B_{\mu\nu}$
can always be written in terms of $g_{\mu\nu}$, since these variables can
of course simply be expressed as the symmetric and antisymmetric
parts of $g_{\mu\nu}$.  One might hope that this objection could be
removed by requiring the theory
to be ``geometrically constructed," a notion still to be defined, but we
shall see that even this constraint turns out not to be sufficient.
A more concrete problem is that the string $\sigma$--model, and
hence the perfectly consistent but non-geometric effective
field theory it generates, is
invariant under the local gauge transformation
\begin{equation}
\delta_0 B_{\mu\nu} = \partial_{\mu} \epsilon_{\nu} - \partial_{\nu}
\epsilon_{\mu} \; ,
\label{Ac}
\end{equation}
in addition to the usual diffeomorphism invariance.
Hence all dependence on $B_{\mu\nu}$ must be through the invariant
field
strength tensor
\begin{equation}
H_{\lambda\mu\nu} \equiv \partial_{\lambda}B_{\mu\nu} +
\partial_\mu B_{\nu\lambda} + \partial_\nu B_{\lambda\mu} \; ,
\label{Ad}
\end{equation}
and indeed the effective string expansion (which is in powers of derivatives,
thereby maintaining consistency)
begins as the sum of Einstein plus $H^2_{\lambda\mu\nu}$ terms.
This puts the effective theory at odds with generic
geometric models constructed from $g_{\mu\nu}$ with the number of
derivatives fixed at two, unless such models
can maintain the above invariance -- which will turn out to be crucial to
their consistency.
Nevertheless, despite these {\it a priori} difficulties of the geometric
approach, it merits reexamination not only in view of the
string motivation, but because we will provide a simple (geometric) extension
of these ideas that restores consistency.

In this paper, we will analyze the generic geometric models,
by expanding them in powers of $B_{\mu\nu}$ about a classical symmetric
background, and consider the merits of the resulting ``gravity $+$ matter''
theories in terms of standard physical criteria:
absence of negative-energy excitations (``ghosts''), of algebraic
inconsistencies, or discontinuities in the degree-of-freedom content.
Our results are negative for the ``geometric'' two-derivative theories.
In brief, our argument is that consistency requires the $B$ expansion to
begin with (quadratic) kinetic terms of $H^2_{\mu\nu\lambda}$ form \cite{KR},
whereas geometric actions homogeneous in two derivatives produce
extensions which necessarily include dependence on powers of undifferentiated
$B_{\mu\nu}$.  These higher-power terms generically violate the
gauge invariance of the leading kinetic
term, with the usual consequences that there are either ghosts,
algebraic inconsistencies or unacceptable constraints on ``independent''
degrees of freedom.  For instance, as we already pointed out in
\cite{DDM1}, NGT suffers from curvature-coupled ghosts.
[Of course the fact that the one consistent, $R + H^2$, two-derivative
model can be formally written as an infinite series of ``geometric''
terms is not to be regarded as a counterexample.]
We will then show that restoration of consistency can nevertheless
be achieved within the framework of geometric theories by adding
``cosmological'', nonderivative terms which render the $B$ field massive,
thereby sidestepping the problems of gauge inconsistencies.
Such extensions will be seen to provide viable alternative gravity
models to be confronted with observation.

In Section 2, we introduce appropriate definitions of geometric theories
constructed from non-symmetric metrics, and the associated expansion
in $B_{\mu\nu}$.  [We also settle there and in Appendix B certain
questions raised in the literature concerning equivalence of
first (Palatini) and second order formulations.]  In Section 3,
we enumerate the generic difficulties faced by these models
and prove that they cannot be overcome by deformation of the abelian gauge
invariance (\ref{Ac}).  Section 4 shows in detail how the two
main previously proposed models fail.
In Section 5, we introduce a new class of consistent geometric
gravity models endowing $B_{\mu\nu}$ with a finite range, couple it to
macroscopic sources and  discuss some possible observational consequences.
All our considerations are intended to be at the purely classical, low
energy level.

In view of the sometimes confusing statements in the literature,
it may be useful to spell out some of our basic assumptions.  The first
of these is that it is meaningful (if not outright mandatory), particularly
at the macroscopic level, to expand the theories in terms of their two
independent coupling constants: the usual gravitational one and a
corresponding one for the $B$-field.  The theory must obviously reduce
to ordinary Einstein gravity when the latter is neglected.
Secondly, we make the physical requirement that, when so
expanded, the theory have no ghost -- negative energy -- modes.
For, even at the classical level, presence of such
modes in the excitation spectrum simply means, if they fail to decouple,
that the universe is liable to be unstable to their radiation.  Indeed, we
will see concretely that, in NGT, such an instability is unavoidable
irrespective of any choice of initial conditions.

\vspace{.2in}
\renewcommand{\theequation}{2.\arabic{equation}}
\setcounter{equation}{0}

\noindent{\bf 2. ~Geometric Theories.}

In the standard tradition, a geometric theory is
defined by an action constructed from generally covariant objects
built with the metric tensor $g_{\mu\nu}$ and the affine connection
$\Gamma^{\lambda}_{\mu\nu}$ on the spacetime manifold.
The  affinity may either be an independent field (first order
formulation) or a prescribed function of $g_{\mu\nu}$ (second order).
These concepts can be largely carried over to models whose
metric tensor and affinity are no longer symmetric.
The Riemann curvature tensor is still
\begin{equation}
R^\mu_{\;\; \nu\alpha\beta} = \partial_{\alpha} \Gamma^\mu_{\nu\beta} -
\Gamma^\sigma_{\nu \alpha} \Gamma^\mu_{\sigma\beta} -
[\alpha \leftarrow\rightarrow \beta] \; ,
\label{Ba}
\end{equation}
but there are now two possible (metric-independent) contractions,
\begin{equation}
R_{\mu\nu} = R^{\lambda}_{\;\; \mu\lambda\nu}\;, \hspace{.5in}
P_{\mu\nu} = R^{\lambda}_{\;\; \lambda\mu\nu}\;.
\label{Bb}
\end{equation}
Note that while $P_{\mu\nu}$ is antisymmetric, $R_{\mu\nu}$ has no
symmetries.  To construct a scalar curvature requires an inverse,
$g^{\mu \nu}$, which we define by the usual convention
\begin{equation}
g^{\mu\alpha}g_{\nu\alpha} = g^{\alpha\mu}g_{\alpha\nu} = \delta^\mu_\nu \;\;
(\neq g^{\mu\alpha}g_{\alpha\nu}) \; .
\label{Bc}
\end{equation}
The antisymmetric part of the connection is the torsion tensor
\begin{equation}
T^\lambda_{\mu\nu} = \textstyle{\frac{1}{2}} \; \Gamma^\lambda_{[\mu\nu ]}
\equiv \textstyle{\frac{1}{2}} \;
\left( \Gamma^\lambda_{\mu\nu} - \Gamma^\lambda_{\nu\mu} \right) \;.
\label{Bd}
\end{equation}
Throughout, we denote unnormalized antisymmetrization/symmetrization
by square/round brackets.

When expanding geometric objects in powers of $B_{\mu\nu}$,
we assume the symmetric part $G_{\mu\nu}$ to be invertible,
and use it to move indices.
Of course, the naive split (\ref{Ab}) can always be composed with a local
algebraic field redefinition, replacing (\ref{Ab}) by
\begin{equation}
g_{\mu\nu} = G_{\mu\nu} + B_{\mu\nu} + \alpha \:
B_{\mu\alpha}\:B^\alpha_{\;\;\nu} +
\beta\: B^{\alpha\beta}B_{\alpha\beta} G_{\mu\nu} + {\cal{O}}(B^3) \;.
\label{Be}
\end{equation}
Note that the generic form (\ref{Be}) incorporates the freedom of defining
the irreducible parts by decomposing a different ``metric" field
variable, say $(\sqrt{-g} )^n g^{\mu\nu}$, so there is no
loss of generality in this respect.  For convenience, we have collected the
relevant expansions in Appendix A.

A geometric theory in first order form is one whose action
depends on $g$ and $\Gamma$ through the curvatures defined above,
as well as on invariants---built from
torsion---which dimensionally respect the second derivative
requirement.
For concreteness, consider NGT which is the simplest generalization
of general relativity in Palatini form,
\begin{equation}
{\cal L}^{(1)}_{\rm NGT}(g,\Gamma) = \sqrt{-g} \: g^{\mu\nu}\:
R_{\mu\nu} (\Gamma ) \; .
\label{Bff}
\end{equation}
In $D$ dimensions, its field equations are
\begin{equation}
(-g)^{-1/2}\;
\partial_\lambda (\sqrt{-g}\:g^{\mu\nu}) +
 \: \Gamma^\mu_{\alpha\lambda}\:g^{\alpha\nu} +
 \: \Gamma^\nu_{\lambda\alpha}\:g^{\mu\alpha} -
 \Gamma^\alpha_{\alpha\lambda}
g^{\mu\nu} -
2 \: \Gamma_\lambda g^{\mu\nu} +
\textstyle{\frac{2}{D-1}}
 \: \: \delta^\nu_\lambda \: \Gamma_\alpha \: g^{\mu\alpha} =  0 \; ,
\label{Bg}
\end{equation}
\begin{equation}
R_{\mu\nu}(\Gamma)~ - ~\textstyle{\frac{1}{2}} \: g_{\mu\nu} \: R ~= ~0
\;,
\label{Bh}
\end{equation}
where
$\Gamma_\mu \equiv \frac{1}{2} \Gamma^\lambda_{[\mu\lambda]}$ and
$R \equiv g^{\mu\nu}R_{\mu\nu}(\Gamma)$.  For $D > 2$, (\ref{Bg})
is equivalent to
\begin{equation}
\partial_\lambda \: g_{\mu\nu} - \widetilde{\Gamma}^\alpha_{\mu\lambda} \:
g_{\alpha\nu} -
\widetilde{\Gamma}^\alpha_{\lambda\nu} \: g_{\mu\alpha} = 0 \; ,
\label{Bi}
\end{equation}
where
\begin{equation}
\widetilde{\Gamma}^\lambda_{\mu\nu} \equiv  \Gamma^\lambda_{\mu\nu} +
\textstyle{\frac{2}{D-1}} \: \delta^\lambda_\mu \: \Gamma_\nu
\label{Bj}
\end{equation}
is constrained by the condition
\begin{equation}
\widetilde{\Gamma}^\lambda_{[\mu\lambda]} = 0 \,.
\label{Bk}
\end{equation}
Combining (\ref{Bj}) and (\ref{Bk}) with (\ref{Bi}) implies
\begin{equation}
\partial_\mu (\sqrt{-g}\: g^{[\mu\nu ]} ) = 0 \; .
\label{Bl}
\end{equation}
  The superficial resemblance of (\ref{Bh}) and (\ref{Bi}) to the usual
Einstein-Palatini system must be viewed with some caution.
One obvious difference is the fact that now just
$\widetilde{\Gamma}^\lambda_{\mu\nu}$,
but not $\Gamma_\mu$, is determined by (\ref{Bi}) in terms of $g_{\mu\nu}$.
Instead, $\Gamma_\mu$ plays the role of a Lagrange multiplier imposing
the constraint (\ref{Bl}).  To see this, note that
\begin{equation}
R_{\mu\nu}(\Gamma ) = R_{\mu\nu}(\widetilde{\Gamma}) -
\textstyle{\frac{2}{D-1}} \: \partial_{[\mu}\Gamma_{\nu]} \; ,
\label{Bm}
\end{equation}
so the Lagrangian ${\cal L}^{(1)}_{NGT}$ splits into
\begin{equation}
{\cal L}^{(1)}_{NGT}(g,\widetilde{\Gamma},\Gamma_\mu) = \sqrt{-g} \:
g^{\mu\nu}\: [ R_{\mu\nu} (\widetilde{\Gamma})
- \textstyle{\frac{2}{D-1}} \partial_{[\mu}\Gamma_{\nu]} ] \; ,
\label{Bn}
\end{equation}
then integrate by parts in the second term of (\ref{Bn}).
[In deriving the field equations from (\ref{Bn}) one
must take account of the constraint (\ref{Bk}).]
A second difference from symmetric theories is that
neither the metric nor the affinity have their familiar
symmetry properties, and thus the index ordering in (\ref{Bi}) is nontrivial.
As was emphasized by
Schr\"{o}dinger, (\ref{Bi}) gives a well-defined
expansion for $\widetilde{\Gamma}^\lambda_{\mu\nu}$
in powers of $B_{\mu\nu}$.  [Actually it has been shown by Tonnelat
\cite{T55}
that (\ref{Bi}) can be solved in closed form; however the resulting expression
is extremely complicated.]  This is by no means the only ordering which does
so since, for example, torsion terms may be added to (\ref{Bi}) as follows
\begin{equation}
\partial_\lambda \: g_{\mu\nu} - \widetilde{\Gamma}^\alpha_{\mu\lambda} \:
g_{\alpha\nu} -
\widetilde{\Gamma}^\alpha_{\lambda\nu} \: g_{\mu\alpha} =
a\widetilde{T}^\alpha_{\mu\lambda}g_{\alpha\nu} +
b\widetilde{T}^\alpha_{\lambda\nu}g_{\mu\alpha} +
c\widetilde{T}^\alpha_{\mu\lambda}g_{\nu\alpha} +
d\widetilde{T}^\alpha_{\lambda\nu}g_{\alpha\mu} \; ,
\label{Bo}
\end{equation}
without altering either its tensorial character or its generic
solvability.  [Indeed the tensorial nature of (\ref{Bi}) is clearest
from the fact that its LHS differs from the usual covariant derivative
of $g_{\mu\nu}$ by
a torsion term.]  There is, however, a line of choices in parameter space,
namely $a+b+c+d+2=0$, for which
(\ref{Bo}) cannot be used to determine $\widetilde{\Gamma}^\lambda_{\mu\nu}$.
We discuss this issue more fully in Appendix A.

Because torsion is available, the above Einstein-Palatini
generalization may be further
extended\footnote {Historically, NGT was selected by
Einstein because it possesses two (physically ill-motivated \cite{Pauli})
symmetries:
transposition invariance ($g_{\mu\nu} \rightarrow g_{\nu\mu}$,
$\widetilde{\Gamma}^\lambda_{\mu\nu} \rightarrow
\widetilde{\Gamma}^\lambda_{\nu\mu}$,
$\Gamma_\mu \rightarrow -\Gamma_\mu$) and ``$\lambda$-invariance''
($\Gamma_\mu \rightarrow \Gamma_\mu + \partial_\mu \lambda$).}
(while maintaining the second derivative requirement) to
\begin{eqnarray}
{\cal L}^{(1)}_{\rm general}(g,\Gamma) & = & \sqrt{-g} \: g^{\mu\nu}
\left[ R_{\mu\nu}(\Gamma ) + a_1P_{\mu\nu}(\Gamma ) + \right.
a_2 \partial_{[\mu} \Gamma_{\nu ]}  \nonumber \\
& + & b_1 \nabla_\lambda T^\lambda_{\mu\nu} + b_2 \:
T^\lambda_{\mu\alpha}T^\alpha_{\lambda\nu} +
b_3 \: T^\lambda_{\mu\nu}\Gamma_\lambda  \nonumber \\
& + & c_1 \: g^{\lambda\delta} g_{\alpha\beta} \:
T^\alpha_{\mu\lambda}T^\beta_{\nu\delta} +
c_2 \: g^{\lambda\delta} g_{\alpha\beta} \:
T^\alpha_{\mu\nu}T^\beta_{\lambda\delta} \nonumber \\
& + & c_3 \: g^{\lambda\delta} g_{\alpha\beta} \:
T^\alpha_{\mu\delta}T^\beta_{\nu\lambda}
\left. + d_1 \: \Gamma_\mu \Gamma_\nu \right] \; .
\label{Bp}
\end{eqnarray}
To understand the nature of the additional terms in light of the
analysis above, let us again decompose $\Gamma^\lambda_{\mu\nu}$ as in
(\ref{Bj}),
to rewrite this as
\begin{eqnarray}
{\cal L}^{(1)}_{\rm general}(g,\widetilde{\Gamma},\Gamma_\mu) & = &
\sqrt{-g} \: g^{\mu\nu}
\left[ R_{\mu\nu}(\widetilde{\Gamma}) + a_1 \: \partial_{[\mu} \right.
\widetilde{\Gamma}^\lambda_{\nu]\lambda} +
b_1 \: \widetilde{\nabla}_\lambda \widetilde{T}^\lambda_{\mu\nu} +
b_2 \: \widetilde{T}^\lambda_{\mu\alpha} \widetilde{T}^\alpha_{\lambda\nu}
\nonumber \\
& + & c_1 \: g^{\lambda\delta}g_{\alpha\beta}
\widetilde{T}^\alpha_{\mu\lambda}\widetilde{T}^\beta_{\nu\delta} +
c_2 \: g^{\lambda\delta}g_{\alpha\beta}
\widetilde{T}^\alpha_{\mu\nu}\widetilde{T}^\beta_{\lambda\delta}
 +  \left. c_3 \: g^{\lambda\delta}g_{\alpha\beta}
\widetilde{T}^\alpha_{\mu\delta}\widetilde{T}^\beta_{\nu\lambda}
\right]
\nonumber \\
& + & \Gamma_\lambda \left[ \left( \textstyle{\frac{2}{D-1}} +
\textstyle{\frac{2D}{D-1}} a_1 - a_2 + \textstyle{\frac{b_1}{D-1}} \right)
\partial_\mu (\sqrt{-g}\:g^{[\mu\lambda]}) \right.
\nonumber \\
& + & \left.  b_3 \widetilde{T}^\lambda_{\mu\nu} \:
\sqrt{-g} \: g^{\mu\nu} + \textstyle{\frac{c_2}{D-1}}
\widetilde{T}^\beta_{\mu\nu} \:
\sqrt{-g} (g_{\alpha\beta}g^{\lambda\alpha} -
g_{\beta\alpha}g^{\alpha\lambda}) \: g^{\mu\nu}
\right. \nonumber \\
& + & \left. \textstyle{\frac{c_3}{D-1}} \widetilde{T}^\beta_{\mu\nu}
(g^{\mu \lambda} g_{\beta \alpha} g^{\alpha \nu} +
g^{\lambda \mu} g_{\alpha \beta} g^{\nu \alpha}) \right] \nonumber \\
& + & \left[ \left( d_1 - \: \textstyle{\frac{b_2}{D-1}} +
\textstyle{\frac{c_2}{(D-1)^2}} +
\textstyle{\frac{2}{D-1}} (c_1+c_3) \right) \sqrt{-g} \:
g^{\mu\nu} \right. \nonumber \\
& - & \left. \textstyle{\frac{c_2}{(D-1)^2}}
\sqrt{-g} g_{\alpha\beta} g^{\mu\alpha} g^{\beta\nu} \right]
\Gamma_\mu\Gamma_\nu \; .
\label{Bq}
\end{eqnarray}
The result is that $\Gamma_\mu$ still imposes a constraint in some sense --
the analogy to gauge-fixing is unmistakable, though as we will
see, misleading.  [Certainly such a gauge-fixing interpretation is
untenable unless
$ b_3 = c_2 = c_3 = 0$, since otherwise
it is a nonlinear ``gauge"
and Faddeev--Popov ghosts would be required.  Accepting this
constraint, the model -- since it contains the multiplier $\Gamma_\mu$ both
linearly and quadratically -- resembles a choice parametrized
between  ``Landau'' and ``Fermi'' type gauges.]  We will not need to use
the full generality in (\ref{Bp}) in
the following; it just serves to show that nothing very dramatic occurs
if we allow it, and that the generalizations (\ref{Bo})
can be obtained from an action.  The models in which
all parameters but $a_1$ and $a_2$ are set to zero have been
especially discussed
in the literature.  It is clear from (\ref{Bq}) that if, further, the
coefficient of the ``$\Gamma_\mu$-constraint" is chosen to
vanish ({\sl i.e.} $2 + 2Da_1 - (D-1)a_2 = 0$), then we have an
exceptional case, but otherwise
these models are all equivalent \cite{Moff} to NGT, Eq. (\ref{Bff}), and stand
or fall with it.

Those members of the general class (\ref{Bp}) for which the affinity cannot
be solved in terms of the metric are not viable geometric
theories and we do not consider them any further.
On the other hand, all other models can of course be expressed in
second order form, {\sl i.e.}, in terms of affinities which are explicit
functions of the metric.  We need therefore only deal with second
order formulations henceforth.
For simplicity, we define $\Gamma(g)$ by what Einstein called the ``$+-$''
relation
\begin{equation}
\partial_\lambda g_{\mu\nu} - \Gamma^\alpha_{\mu\lambda}g_{\alpha\nu} -
\Gamma^\alpha_{\lambda\nu}
g_{\mu\alpha} = 0 \; .
\label{Br}
\end{equation}
Second order geometric theories are then defined by the requirement
that the action depend on $g_{\mu\nu}$ (and its inverses) with all
derivatives appearing in general covariants constructed from
$\Gamma^\lambda_{\mu\nu}(g)$ defined through (\ref{Br}).

As for all theories involving torsion, there is of course a difference
between the
first order theory (\ref{Bff}) and the naive second order theory defined
by the same Lagrangian with $\Gamma^{\lambda}_{\mu\nu} =
\Gamma^{\lambda}_{\mu\nu}(g)$ determined by (\ref{Br})
(see (\ref{Bv}) below).
{}From the discussion above it is clear that to
obtain the equivalent second order formulation of NGT a Lagrange
multiplier field $b_\mu$ must be introduced, and in fact the
model equivalent to (\ref{Bff}) is
\begin{equation}
{\cal L}^{(2)}_{NGT}(g,b) = \sqrt{-g} \: g^{\mu\nu} R_{\mu\nu}(\Gamma (g)) -
b_\nu \: \partial_\mu (\sqrt{-g} \; g^{[\mu\nu ]}) \; .
\label{Bs}
\end{equation}
For, upon varying (\ref{Bs}), we obtain
\begin{equation}
R_{\mu\nu}  + \partial_{[ \mu} b_{\nu ]} = 0 \;, \;\;\;\;
\partial_\mu (\sqrt{-g}\: g^{[\mu\nu ]} ) = 0 \; ,
\label{Bt}
\end{equation}
from which we learn that
\begin{equation}
\Gamma^\lambda_{\mu\nu}(g) = \widetilde{\Gamma}^\lambda_{\mu\nu}(g) \; .
\label{Bu}
\end{equation}
Thus the field equations are equivalent to those of the
original first order system, the Lagrange multiplier
$b_\mu$ now taking the role formerly played by $\Gamma_\mu$
(more precisely, $b_\mu = \textstyle{\frac{-2}{D-1}} \Gamma_\mu$).
On the other hand, the simplest second-order theory, defined by
Eq. (\ref{Bs}) without the Lagrange multiplier
(and introduced \cite{Mann} under the name ``Algebraically extended
Hilbert Gravity'' (AHG)), is
\begin{equation}
{\cal L}^{(2)}_{AHG}(g) = \sqrt{-g} \: g^{\mu\nu} R_{\mu\nu}(\Gamma (g)) \, ,
\label{Bv}
\end{equation}
with $\Gamma^\lambda_{\mu\nu}(g)$ determined by (\ref{Br}), and is very
different
from NGT.
The consistency of both NGT and AHG will be critically examined below in
Section 4.

  By this point, the reader may have noted some pedantry in our
introduction
of the geometric theories and their equivalent formulations.  Our
motivation
for this development was twofold.  Firstly, to the extent that we will be
obtaining no-go theorems, it is important to establish precisely what
does
not go.  Secondly, the literature contains occasional confusing claims,
for example about inequivalence of first and second order
formulations, so it also seems relevant to state exactly what we believe.
At the risk of overkill, we will even give an explicit derivation of
equivalence
-- for a model where it has been claimed otherwise, namely the
linearizations of (\ref{Bff}) and (\ref{Bs}) -- in Appendix B.

To conclude this ``kinematical" section, we return to the fact that with a
looser definition of ``geometrical," one can write any generally covariant
action involving the standard symmetric metric $G_{\mu\nu}$ and the
matter field $B_{\mu\nu}$ in geometrical form: given any desired
term, we may simply identify $G_{\mu\nu}$ and $B_{\mu\nu}$ as the symmetric
and antisymmetric pieces of $g_{\mu\nu}$ respectively.
Then one is left with the problem of
representing the inverse $G^{\mu\nu}$ in
terms of $g_{\mu\nu}$ and {\it its} inverse
$g^{\mu\nu}$.  Of course the penalty which one must pay is that
an infinite number of terms will now be required in general to obtain this
representation, so this constitutes far too
loose a way to ``unify" the fields!  A somewhat more geometric
rewriting of a generic term seems possible if one uses the fact that
$g^{\mu\alpha}g_{\alpha\nu} \neq \delta^\mu_\nu $, which
implies that a given string of terms built
from the metric and its inverse -- {\sl e.g.},
$g^{\mu\alpha}g_{\alpha\beta}g^{\beta\gamma}g_{\gamma\nu}\:...$, with
indices distributed
as shown -- will be nontrivial.  [Indeed, a counting of parameters -- at
least to fourth order in the expansion -- suggests
that these strings may be used in combinations of curvature and torsion
invariants to reproduce the desired term to any given order.]

\vspace{.1in}
\noindent{\bf 3. ~Expansion about a symmetric background.}

  In order to understand the consistency problems faced by generic
geometric models, it is simplest to expand the field equations in powers of
the $B_{\mu\nu}$ field about a general curved symmetric background
$G_{\mu\nu}$.
Because there has been some confusion about expandability
in the literature (see, {\it e.g.}, \cite{CoMof}),
we emphasize that our analysis is based on
the fact that the dynamics of these two separate components,
$G_{\mu\nu}$ and $B_{\mu\nu}$, can be described
by two independent coupling constants -- namely the usual Einstein $\kappa^2$
and that associated with the antisymmetric field.  We assume (as is
standard in field theory) that expansion in these coupling constants
is allowed, so that in particular the theory must reduce to Einstein gravity
to zeroth order in the antisymmetric field, and require that it remain
consistent order by order (without discontinuities in the degree-of-freedom
content).  [For example, the generalization of the Schwarzschild solution
in NGT is separately analytic in these two parameters and satisfies
(\ref{De} -- \ref{Dg}) below to the expected order in $B$.]
Note that in general
relativity there is a rigorous proof that the expansion in $\kappa^2$
is perfectly legal -- even for studying the asymptotic behaviour at infinity --
in that it generates a series which is the Taylor expansion of some exact
solution of the theory \cite{TDexp}.

  We will divide our analysis into two parts, according to whether the field
equations are homogeneous of second derivative order, or also include
nonderivative -- ``cosmological'' -- terms.  Henceforth we
work in second-order formalism only, omitting
the corresponding superscript 2 on Lagrangians.

\noindent a)  ~\underline{No ``cosmological'' terms}

   The leading terms to consider in the action are those quadratic
in $\partial B$.  It is well known that (just as in electrodynamics) the
only ghost-free action quadratic in a massless $B$ field is the square of
its field strength, (\ref{Ad}), whatever the gravitational background:
\renewcommand{\theequation}{3.\arabic{equation}}
\setcounter{equation}{0}
\begin{equation}
{\cal L}_2 = - \textstyle{\frac{1}{12}} \sqrt{-G} \;
H_{\mu\nu\lambda}\: H^{\mu\nu\lambda} \; ,
\label{Ca}
\end{equation}
which is gauge invariant under (\ref{Ac}).  [We recall that all indices are
moved by $G_{\mu\nu}$, our signature is mostly plus, and
that the cyclic ordinary derivative (``curl") in the field strength (\ref{Ad})
defines a tensor.]
Hence any geometric action inequivalent (modulo field redefinitions)
to (\ref{Ca}) at order $(\partial B)^2$ already contains ghosts, and
so cannot provide an acceptable effective action.
[We shall see shortly that this conclusion holds even in the presence of
apparent ``gauge-fixing'' terms, as in (\ref{Bs}) for NGT.]
The requirement (\ref{Ca}), already necessary in flat space, is however
far from sufficient when working about generic backgrounds.
Consider, for example, a typical term in a geometric action (from Appendix A),
\begin{eqnarray}
\sqrt{-g} \: g^{\mu\nu}R_{\mu\nu}(g) & = &
\sqrt{-G} \: \bar{R} (G) - \textstyle{\frac{1}{12}} \:
\sqrt{-G} \: H^{\mu\nu\lambda}H_{\mu\nu\lambda} +
\textstyle{\frac{1}{2}} (\textstyle{\frac{1}{2}} - \alpha + (D-2)\beta)
\sqrt{-G} \: \bar{R}(G)B^{2} \nonumber \\[.1in]
& - & \alpha \sqrt{-G} \: \bar{R}_{\mu\nu} (G)B^{\mu\alpha}B_\alpha^{\;\;\nu}
- \sqrt{-G} \: \bar{R}_{\mu\nu\alpha\beta} (G) \: B^{\mu\alpha}B^{\nu\beta}
+ {\cal O}(B^4) \, ,
\label{Cb}
\end{eqnarray}
where bars denote covariantization with respect to the $G_{\mu\nu}$
background and we have dropped total derivative terms.
In addition to the correct $(\partial B)^2$ terms (\ref{Cb})
also contains (dimensionally equivalent) background-curvature-coupled
terms $\sim \bar{R}BB$.  The latter fall into two classes:
those involving only the Ricci tensor (or scalar) are removable by field
redefinition, with the appropriate choice of $\alpha ,\beta$ in (\ref{Be}).
However, terms involving the full curvature cannot be so removed
(for $D > 3$).  Their
presence is fatal to the consistency of the theory because they destroy
gauge invariance, which leads to unacceptable local constraints.
This will be elaborated for the key cases in Section 4.
It is clear from the formulas of Appendix A, that the consistency requirements
at quadratic order in $B$ rule out almost all the second order actions
of the form (\ref{Bp}).
It is nevertheless possible to achieve a consistent action at quadratic
order (pure $H^{2}$ without $\bar{R}BB$ terms) by combining several
``geometric" terms.  Using the results of Appendix A, an explicit example
of a quadratically acceptable action is
\begin{equation}
{\cal L} = \sqrt{-g} g^{\mu\nu} [ R_{\mu\nu}(g) - \Gamma_\mu\Gamma_\nu -
T^\alpha_{\mu\lambda}T^\lambda_{\alpha\nu} - \nabla_\alpha T^\alpha_{\mu \nu}
- g_{\alpha \beta} g^{\rho \sigma} T^\alpha_{\mu \rho} T^\beta_{\nu \sigma} ]
\label{Cc}
\end{equation}
after appropriate choice of field redefinition.
Let us consider now the general class
of quadratically acceptable actions,
\begin{equation}
{\cal L}_d = {\cal L}_2 + \sum_{n \geq 3} \:
{\cal L}_n \equiv {\cal L}_2 + {\cal L}_> \; .
\label{Cd}
\end{equation}
Here ${\cal L}_d$ is a modification of the required
gauge invariant ${\cal L}_2$ of (\ref{Ca}) by higher order
$(n > 2)$ terms in $B_{\mu\nu}$ (with 2 derivatives or a curvature term).
Clearly, an expression of the
form (\ref{Cd}) is also not generally consistent since it involves terms
of the form $\bar{R}B^n$ or $\partial B \partial B B^{n-2}$, so that it loses
the initial invariance $(\delta_0 )$ of $H_{\mu\nu\lambda}$.  This
loss of the local invariance entails
unacceptable conservation constraints (notorious in higher
spin gauge theories) on the nonlinear sources of the {\it
identically conserved} linear term in the field equations.
So, these models are only saved if the abelian invariance itself can be
deformed to
\begin{equation}
\delta = \delta_0 + \sum_{n\geq 1} \delta_n \equiv \delta_0 + \delta_>
\label{Ce}
\end{equation}
such that the action $S_d$ is invariant under the deformed transformation.
If such a deformation exists at all, it will also be present in particular
when
the background is flat, $G_{\mu\nu} = \eta_{\mu\nu}$.  [The only exception
would be a deformation of the full theory which reduced to the
linear invariance at $G_{\mu\nu} = \eta_{\mu\nu}$.  However, the
deformation
$\delta_>$ must then be built with curvatures, and thus would contain
too
many derivatives to produce an invariance of a two-derivative order
action.]
We now show that there is no such deformation available even
about a flat background (for an initial attempt at finding such
a deformation, see the third reference cited in \cite{Mann}).

   We preface a demonstration of the nonexistence of deformations of
(\ref{Ca}) based on the technical approach of \cite{Wald}, by giving a
transparent (though somewhat heuristic) argument for the case of
Maxwell theory.  In that case, one may ask for a consistent deformation of
${\cal L}^M(\partial A) = -\textstyle{\frac{1}{4}}\: F^{2}_{\mu\nu}$ by
two-derivative terms ${\cal L}^M_>$ of higher
order in $A_\mu$, invariant under the extension $\delta = \delta_0 \: +
\delta_>$ with
$\delta_0 A_\mu = \partial_\mu \epsilon$.  This was shown not to exist
in \cite{Wald}.  Another way to see the
problem is as follows.  Let ${\cal L}_{n_{0}} , \; n_0 > 2$, be the
first nonvanishing deformation; we then need $\delta_{n_0-2}$ with
\begin{equation}
\delta_0 S^M_{n_{0}} = -\delta_{n_{0}-2} \: S^M_2  \; .
\label{Cf}
\end{equation}
But then
\begin{equation}
\left. \partial_\mu \; \frac{\delta S^M_{n_{0}}}{\delta A_\mu} \right|_0
\equiv \partial_\mu J^\mu = 0
\label{Cg}
\end{equation}
where $|_0$ means evaluated on
free $(S^M_2 )$ mass shell.  Now if
$J^\mu$ is a conserved current on
linearized shell, there must (by Noether's theorem) be a corresponding
continuous symmetry (with a scalar parameter $\widetilde{\epsilon}$ say) of
the
Maxwell action, local of order $n_0-2$ in $A_\mu$ and of first derivative
order.  [This ignores identically conserved ``superpotential" currents
$K^\mu \equiv \partial_\nu Z^{[\mu\nu]}$
which should be removed by field redefinitions.]
Basically the statement is that there are none such -- and this may be
verified by explicit enumeration (if nothing else!) --
except those equivalent to the original local gauge invariance,
\begin{equation}
\widetilde{\delta}A_\mu = \widetilde{\epsilon} \: \partial_\mu f(A) \; ,
\label{Ch}
\end{equation}
whose Noether current is
\begin{equation}
J^\mu = (\partial_\nu f(A)) F^{\mu\nu}(A) \; .
\label{Ci}
\end{equation}
This is indeed a conserved current on linear shell, and
in fact the field equation
$\partial^\nu F_{\mu\nu} = (\partial^\nu f(A)) F_{\mu\nu}$ is perfectly
consistent, though not gauge invariant.
However, it cannot be derived from an action; thus one cannot find any
consistent deformations of the Maxwell action and its gauge invariance.
This argument may probably be extended to the $B$-field, but
for technical ease we will follow \cite{Wald}.
  Going back to our $B_{\mu\nu}$ problem, (\ref{Cd}), let us begin
with a quick review of the main points in the
approach of \cite{Wald}.  [Note that we do not derive here the most
general result possible, but simply demonstrate it in the context of actions
homogeneous in two derivatives as is relevant to our discussion.]
Denote the equation of motion by ${\cal F}^{\mu\nu}$,
\begin{equation}
{\cal F}^{\mu\nu} = \frac{\delta S_d}{\delta B_{\mu\nu}} = {\cal F}^{\mu\nu}_1
+ {\cal F}^{\mu\nu}_2 + \; ... \;\; ,
\label{Cj}
\end{equation}
where ${\cal F}^{\mu\nu}_1 = \textstyle{\frac{1}{2}}
\partial_\lambda H^{\lambda\mu\nu}$, and
the ``linearized Bianchi identity" is
\begin{equation}
\partial_\mu {\cal F}^{\mu\nu}_1 \equiv 0  \; .
\label{Cl}
\end{equation}
For the deformed theory to preserve its degree-of-freedom content, the field
equations (\ref{Cj}) must admit a corresponding (nonlinear) Bianchi identity,
which can be written as
\begin{eqnarray}
\partial_\mu(U_{\alpha\beta}^{\mu\nu}{\cal F}^{\alpha\beta}) & = &
U_{\alpha\beta}^{\mu\lambda}V^\nu_{\mu\lambda}{\cal F}^{\alpha\beta} \; ,
\nonumber \\
U_{(0)\alpha\beta}^{\;\;\mu\nu} & = & \delta^\mu_{[\alpha} \delta^\nu_{\beta]}
\;\; , \;\;
V_{(0)\alpha\beta}^{\;\;\lambda} = 0 \; .
\label{Cm}
\end{eqnarray}
In turn, the identity (\ref{Cm}) is equivalent to requiring invariance of the
action $S_d$ under the deformed gauge transformation
\begin{equation}
\delta B_{\mu\nu} = U^{\alpha\beta}_{\mu\nu}(\partial_\alpha \epsilon_\beta+
V^\lambda_{\alpha\beta} \epsilon_\lambda)
\; .
\label{Cn}
\end{equation}
Here, consistent with the derivative constraints,
$U_{\alpha\beta}^{\mu\nu} = -U_{\beta\alpha}^{\mu\nu}$ is algebraic in
$B_{\mu\nu}$
and $V_{\alpha\beta}^\lambda$ contains at most one derivative of $B_{\mu\nu}$.
Clearly this is a redundant parametrization, in that $(U,V)$
is equivalent to the set $(U',V')$ for which there is a matrix $f$
algebraic in $B_{\mu\nu}$, $f_{(0)\mu}^{\;\;\nu} = \delta^\nu_\mu$,  such that
\begin{eqnarray}
(U')^{\mu\nu}_{\alpha\beta} & = & f^\nu_\gamma U_{\alpha\beta}^{\mu\gamma}
\nonumber \\
(V')^\lambda_{\alpha\beta} & = & f^\lambda_\delta V^\delta_{\alpha\gamma}
(f^{-1})^\gamma_\beta+
(f^{-1})_\beta^\gamma\partial_\alpha f^\lambda_\gamma\; .
\label{Co}
\end{eqnarray}
Further, one easily sees that a field redefinition only affects $U$,
and that there exists (locally) a field redefinition which sets
$U_{\alpha\beta}^{\mu\nu} = \delta^\mu_{[\alpha} \delta^\nu_{\beta]}$ if and
only if
\begin{equation}
\frac{\partial U_{\alpha\beta}^{\mu\nu} }{\partial B_{\gamma\delta}}
U^{\rho\sigma}_{\gamma\delta} -
\frac{\partial U_{\alpha\beta}^{\rho\sigma} }{\partial B_{\gamma\delta}}
U^{\mu\nu}_{\gamma\delta} = 0
\; .
\label{Cp}
\end{equation}
One obtains constraints on the existence of a deformed gauge invariance
of the form (\ref{Cn}) by demanding the closure of the gauge algebra,
namely that it be possible to find a transformation parameter
$\epsilon''$ (depending on $\epsilon, \epsilon' , B$) such that
\begin{equation}
[\delta,\delta'] = \delta'' \; .
\label{Cq}
\end{equation}
The strategy is to try to determine $U$ and $V$ order by order in
$B_{\mu\nu}$,
by imposing (\ref{Cq}).   At the first order we have candidates
(the symmetric part of $V$ at this order will be seen to be irrelevant)
\begin{eqnarray}
U_{(1)\alpha\beta}^{\;\;\mu\nu} & = &
b_1B_{[\alpha}^{\;\;\mu}\delta_{\beta]}^\nu +
b_2B_{[\alpha}^{\;\;\nu}\delta_{\beta]}^\mu \nonumber \\
V_{(1)[\alpha\beta]}^{\;\;\lambda} & = & a_1 \partial^\lambda B_{\alpha\beta}
+ a_2 \partial_{[\alpha}B_{\beta]}^{\;\; \lambda} +
a_3 \partial^\gamma B_{\gamma[\alpha} \delta_{\beta]}^\lambda \; .
\label{Cr}
\end{eqnarray}
By using the freedom (\ref{Co}), with $f_{(1)\mu}^{\;\;\nu} \sim
B_\mu^{\;\;\nu}$,
we may set $b_1=-b_2$.   To zeroth order these are all that are
required
in (\ref{Cq}), for which at this order the r.h.s. is just
$\partial_{[\mu}\epsilon''_{(0)\nu]}$.
Thus we may simplify by taking $\partial_\lambda$ and totally antisymmetrizing
over $[\lambda\mu\nu]$, to obtain
\begin{eqnarray}
& - & 2b_2[\partial_{[\mu}
\epsilon^\alpha\partial_\lambda\partial_\alpha\epsilon'_{\nu]} -
\partial_{[\lambda}\partial^\alpha\epsilon_\mu\partial_\alpha\epsilon'_{\nu]}
- (\epsilon \leftarrow\rightarrow \epsilon')]
+ (a_1-a_2)[\partial^\alpha\partial_{[\mu}\epsilon_\nu\partial_{\lambda]}
\epsilon'_\alpha -
(\epsilon \leftarrow\rightarrow \epsilon')] + \nonumber \\
& + & a_3[\Box \partial_{[\lambda}\epsilon_\mu \epsilon'_{\nu]} +
\Box\epsilon_{[\mu}\partial_\lambda\epsilon'_{\nu]} -
\partial_{[\mu}\partial^\alpha\epsilon_\alpha\partial_\lambda\epsilon'_{\nu]} -
(\epsilon \leftarrow\rightarrow \epsilon')] = 0 \; .
\label{Cs}
\end{eqnarray}
Setting $\epsilon_\mu$ to be a constant vector we learn that
$a_3 \Box \partial_{[\lambda}\epsilon'_\mu \epsilon_{\nu]} = 0$ for all
$\epsilon'$, and
thus $a_3=0$.  Similarly, taking $\partial_\mu\epsilon_\nu = S_{\mu\nu} =
S_{\nu\mu}$
constant, and $\partial_\mu\epsilon_\nu = A_{\mu\nu} = -A_{\nu\mu}$
constant, we
find
$b_2=0$ and $a_1=a_2$.  Thus imposing the closure constraint (\ref{Cq}) to
zeroth order in $B$ has reduced the first order deformation to
\begin{eqnarray}
U_{(1)\alpha\beta}^{\;\;\mu\nu} & = & 0 \nonumber \\
V_{(1)[\alpha\beta]}^{\;\;\lambda} & = & a_1 H^\lambda_{\;\;\alpha\beta} \; .
\label{Ct}
\end{eqnarray}
Moreover, inserting (\ref{Ct}) into (\ref{Cq}) yields
$\partial_{[\mu}\epsilon''_{(0)\nu]} = 0$, from which we deduce that
$\epsilon''_{(0)\mu} = 0$ because it is impossible to construct a
pure gradient which is bilinear in $\epsilon_\mu$ and $\epsilon'_\mu$
and antisymmetric under their exchange.
If we now consider the special case for which
$\epsilon_\mu = \partial_\mu \epsilon$ and
$\epsilon'_\mu = \partial_\mu \epsilon'$, then (\ref{Ct}) is all
we require to impose (\ref{Cq}) to {\it first} order since the zeroth
order variation of $B$ vanishes.
The r.h.s. of (\ref{Cq}) then reduces at
lowest order to $\partial_{[\mu}\epsilon^{''}_{(1)\nu]}$, where
$\epsilon^{''}_{(1)\nu}$ is linear in $B$.  As above, on taking the
curl of the resulting equation we obtain a simple constraint
which can generically only be satisfied if $a_1$ vanishes.
Another way to prove that necessarily $a_1=0$, is to check that there is
no combination of the two possible nontrivial terms in a cubic action,
\begin{equation}
{\cal L}_3 = B_{\mu\nu} \left[
A_1 \partial^\mu B^{\nu\beta}\partial^\alpha B_{\alpha\beta} +
A_2 \partial^\mu B_{\alpha\beta}\partial^\alpha B^{\nu\beta} \right] \; ,
\label{Cv}
\end{equation}
which can compensate the change of (\ref{Ca}) under
$\delta_1B_{\mu\nu} = H^{\lambda}_{\;\;\mu\nu}\epsilon_\lambda$,
{\it i.e.}, which solves $\delta_1{\cal L}_2 = - \delta_0{\cal L}_3$.
Having proven the triviality of all possible first order deformations,
an inductive argument using (\ref{Cq}) exactly
as in \cite{Wald} shows that
\begin{eqnarray}
U_{(n)\alpha\beta}^{\;\;\mu\nu} & = & 0 \;\; , n \geq 1 \nonumber \\
V_{(n)[\alpha\beta]}^{\;\;\lambda} & = & 0 \;\; , n \geq 0 \; .
\label{Cu}
\end{eqnarray}
Hence we have shown that it is not
possible to produce a deformation which will give a consistent
generalization of the gauge invariance in this case.  As the only
gauge invariant
action\footnote{Note however the possibility, when $D=5$, of
Chern-Simons-type kinetic terms:
$\int{d^5x} \epsilon^{\mu\nu\lambda\alpha\beta} H_{\mu\nu\lambda}
B_{\alpha\beta}$. }
that can be written down in $D=4$ is
$\int{d^4x} \sqrt{-G} H^2$, we conclude that (\ref{Ca}) defines
the only consistent theory in the absence of cosmological
terms.

  We now briefly return to close the remaining loophole, that perhaps
some terms beyond (\ref{Ca}) appearing at quadratic order could be
interpreted as ``gauge-fixing".  The main point here is that the very
notion of gauge-fixing does not make sense unless there is an underlying gauge
invariant theory -- and as we have just shown, there is no natural
candidate in this geometric context.  Therefore all the Lagrangians (\ref{Bp})
containing $\Gamma_\mu$ multipliers will necessarily exhibit propagating
ghosts, without the possibility of projecting onto
an invariant ghost-free subspace.  It is here that  the lack of underlying
gauge invariance is lethal -- this moral is well known, and we
content ourselves with illustrating it in an example in Section 4.

\noindent b) ~\underline{With `Cosmological' terms}

Suppose we now also permit ``cosmological terms" ({\sl i.e.}, terms
without derivatives) in the Lagrangian.  The archetypal example is
$\sqrt{-g} = \sqrt{-G} (1 + \frac{1}{4} B^{2} + ...)$, but
we could also arrange to cancel $\sqrt{-G}$ or in general give different
coefficients to $\sqrt{-G}$ and the $B^{2}$ term (see Appendix A).
Perhaps
surprisingly, this alters the no-go conclusions of the previous
subsection despite the fact that $\sqrt{-g}$ is certainly geometric
in the narrowest sense.  [As emphasized by Schr\"{o}dinger, who
favoured them \cite{ES}, such models can even be obtained from the
geometrically elegant -- purely affine -- Lagrangian
$\sqrt{-\rm{det} R_{\mu\nu}(\Gamma)}$.]  The point is
that once a mass term ($\sim -\frac{1}{4} m^2 B^2$)
is included, then the obvious constraints associated with gauge invariance
disappear -- for example we do not need to attempt the
(fruitless) deformation route to avoid conservation inconsistencies.
[Of course, the second derivative quadratic term
must still be $H^{2}$ for ghost-freedom, just as it is only
$-\frac{1}{4} \: F^{2} - \frac{1}{2} m^2 A^2$, but
not $\frac{1}{2} A^\mu (\Box - m^{2})A_\mu$
which is an acceptable finite range vector Lagrangian.]  As an
illustrative
example, consider the Lagrangian
\begin{equation}
{\cal L} = - \textstyle{\frac{1}{12}}(1 + f_1B^2)H^2_{\lambda\mu\nu} -
\textstyle{\frac{1}{4}}m^2B^2 - f_2B^2B^2 \;,
\label{Cw}
\end{equation}
where $f_i$ are some coupling constants.  The equations of motion
are
\begin{equation}
\textstyle{\frac{1}{2}}\partial_\lambda [(1+f_1B^2)H^{\lambda\mu\nu}] -
\textstyle{\frac{1}{6}}f_1H^2B^{\mu\nu} -
\textstyle{\frac{1}{2}}m^2B^{\mu\nu} - 4f_2 B^2B^{\mu\nu} = 0 \; .
\label{Cx}
\end{equation}
We may look for a solution expanded in ``powers of nonlinearity'',
$B_{\mu\nu} = \sum_{n \geq 0}B_{\mu\nu}^{(n)}$ around $B^{(0)} = 0$.
Up to third order, the equations are
\begin{eqnarray}
\textstyle{\frac{1}{2}}\partial_\lambda H^{\lambda\mu\nu}(B^{(1)}) & - &
\textstyle{\frac{1}{2}} m^2 (B^{(1)})^{\mu\nu}  =  0 \nonumber \\
\textstyle{\frac{1}{2}}\partial_\lambda H^{\lambda\mu\nu}(B^{(2)}) & - &
\textstyle{\frac{1}{2}} m^2 (B^{(2)})^{\mu\nu}  =  0
\label{Cy}
\\
\textstyle{\frac{1}{2}}\partial_\lambda H^{\lambda\mu\nu}(B^{(3)}) -
\textstyle{\frac{1}{2}} m^2 (B^{(3)})^{\mu\nu}
& + & \textstyle{\frac{1}{2}}\partial_\lambda[f_1(B^{(1)})^2H^{\lambda\mu\nu}
(B^{(1)})] - 4f_2(B^{(1)})^2(B^{(1)})^{\mu\nu} \nonumber \\
& - & \textstyle{\frac{1}{6}}f_1 (H(B^{(1)}))^2 B^{(1) \mu \nu}
=  0 \, . \nonumber
\end{eqnarray}
As long as $m^2 \neq 0$ there is no problem with consistency
constraints
from lower order equations on the higher order ones.   For example,
now
using the identity (\ref{Cl}) simply determines that
$\partial^\mu B^{(1)}_{\mu\nu} = 0 = \partial^\mu B^{(2)}_{\mu\nu}$, and
gives a relation for $\partial^\mu B^{(3)}_{\mu\nu}$ in terms of the
lower order terms in the solution.

  Although $m^2 \neq 0$ avoids the local consistency constraints
in a perturbative expansion in $B_{\mu\nu}$, more work would be
needed to check whether, from a mathematical standpoint, the field
equations associated with say (\ref{Cw}) form a well-posed problem,
with acceptable causality properties.

\vspace{.2in}

\noindent {\bf 4. ~Problems with the basic models.}

In light of our discussion in Section 3a, it should be clear that a generic
(massless) geometric model will violate either ghost-freedom or local
consistency requirements.  In this section, we will illustrate these
failings by studying in some detail the two main
examples of geometric models which have been proposed in the literature.

Consider first the simplest second order theory, (AHG) \cite{Mann},
\renewcommand{\theequation}{4.\arabic{equation}}
\setcounter{equation}{0}
\begin{equation}
{\cal L}^{(2)}_{AHG}(g) = \sqrt{-g} \: g^{\mu\nu} \:R_{\mu\nu} (g) \; .
\label{Da}
\end{equation}
As we saw from (\ref{Cb}), this model's expansion already has a term
quadratic
in $B_{\mu\nu}$, proportional to
the full Riemann tensor. It can neither be
field redefined away, nor is it $\delta_0$--gauge
invariant.  Thus the model is sick already at quadratic order, having
``ghost" kinematics, even though its flat space limit is the required
$H^{2}$ theory.  This much was noticed in \cite{Kelly}; earlier it had
separately been noted that these theories had no
asymptotically flat static spherically symmetric solution \cite{KellMann}.
To elucidate further the sickness of the model when expanded on a curved
background, let us consider the field equations, which decompose into
symmetric and antisymmetric pieces as follows:
\begin{eqnarray}
\bar{R}_{\mu\nu} + {\cal{O}}(B^{2}) & = & 0 \nonumber \\
\bar{\nabla}^\alpha \: H_{\alpha\mu\nu} -
4 \bar{R}_{\mu\alpha\nu\beta}B^{\alpha\beta} +
{\cal{O}}(B^3) & = & 0 \; .
\label{Db}
\end{eqnarray}
As we are only looking for a family of solutions continuously connected
to those
of general relativity, we can ignore
the terms of higher order in $B$ in these equations.
By the identical conservation of $\bar{\nabla} H$, we then have
the consistency requirement
\begin{equation}
\bar{\nabla}^\mu \bar{R}_{\mu\alpha\nu\beta}B^{\alpha\beta} +
\bar{R}_{\mu\alpha\nu\beta} \bar{\nabla}^\mu B^{\alpha\beta} = 0 \; ,
\label{Dc}
\end{equation}
which by the Bianchi identities (using $\bar{R}_{\mu\nu}=0$) can be
simplified to
\begin{equation}
\bar{R}_{\mu\alpha\nu\beta} \bar{\nabla}^\mu B^{\alpha\beta} = 0 \; .
\label{Dd}
\end{equation}
This is an unacceptable local constraint on the ``independent" degrees
of freedom.

Our second example is NGT, which has been regarded as the remaining
geometric model apparently not disqualified by previous work.
Its linearization about flat space, considered
in Appendix B, consists of linearized Einstein theory plus a gauge fixed
version of the $H^{2}$ action \cite{Moff}.  One might therefore have
hoped that its undesirable excitations could always be
projected away in a consistent fashion, and thus it would exemplify ``geometric
theories with built-in gauge fixing," as discussed at the end of section
3a.  We now show, however, that NGT is inconsistent at the next level
where the $G$-background is no longer flat.  Thus the analogy to gauge
fixing will be seen to be invalid, consistent with our previous
general conclusions.  [The following discussion
amplifies on the result briefly announced in \cite{DDM1},
and is included for completeness.]

Using (\ref{Fh}), the field equations, to zeroth and linear order in
$B_{\mu\nu}$,
become
\begin{eqnarray}
\bar{R}_{\mu\nu} & = & 0  \label{De} \\
\bar{\nabla}^\lambda H_{\lambda\mu\nu} -
4\bar{R}_{\mu\;\;\nu}^{\;\;\alpha\;\;\beta} B_{\alpha\beta}
& = & \textstyle{\frac{4}{D-1}} \partial_{[\mu} \Gamma_{\nu ]} \label{Df} \\
\bar{\nabla}^\mu B_{\mu\nu} & = & 0 \label{Dg}
\end{eqnarray}
upon inserting the solution for $\widetilde{\Gamma}$ to
this order into the $g_{\mu\nu}$ field equations.  In
a flat background, this system of course reduces to the set
(\ref{Ge}), (\ref{Gf}), which is shown
there to be precisely a gauge fixed system.  The gauge condition
(\ref{Gf}) still admits a residual gauge invariance, namely
$\delta B_{\mu\nu} = \partial_{[\mu} \epsilon_{\nu ]}$ with
$\partial^\mu \partial_{[\mu} \epsilon_{\nu ]} = 0$.
If we consider instead a generic Ricci-flat background (\ref{De}),
then the putative residual gauge invariance
\begin{equation}
\delta B_{\mu\nu} = \partial_{[\mu} \epsilon_{\nu ]} \; , \;\;\;\;
\bar{\nabla}^\mu \bar{\nabla}_{[\mu} \epsilon_{\nu ]} = 0
\label{Dk}
\end{equation}
is clearly lost because of the $\bar{R} B$ term in Eq. (\ref{Df}).
This is an immediate indication that the system in a generic background
is no longer the gauge-fixed version of a ghost-free theory.  [A contrary
statement in \cite{Kelly} has since been corrected in \cite{Kerr}.]

  A more direct way to exhibit the problems of the system
(\ref{De}) -- (\ref{Dg}) consists of taking the divergence
of (\ref{Df}).  This yields
\begin{equation}
\bar{\nabla}^\mu \bar{\nabla}_{[\mu}\Gamma_{\nu ]} =
-(D-1) \bar{R}_{\mu\;\;\nu}^{\;\;\alpha\;\;\beta} \; \bar{\nabla}^\mu \:
B_{\alpha\beta} \,,
\label{Dh}
\end{equation}
which is an inhomogeneous Maxwell equation for $\Gamma_\mu$.
Consequently, one can no longer fix initial conditions to eliminate
$\Gamma_\mu$ invariantly, and these dangerous
extra modes do not decouple.

  We now show how the existence of a propagating $\Gamma_\mu$
implies the presence of ghost (negative energy) modes.  To be explicit, let
us first summarize how energy is to be calculated
in the theory, or more relevantly, in our expansion scheme.  For
simplicity, we consider the problem in $D=4$ spacetime and
we treat the case in which there is no normal cosmological term $\sqrt{-G}$;
its presence would not alter our conclusions.  Then, we
require asymptotic flatness, and hence sufficiently rapid decay of the
$B$ field at infinity.  This permits us to split the action through
order $B^2$ into ``Einstein plus matter'' form:
\begin{eqnarray}
I & = & \int{d^4x} \sqrt{-G} [\bar{R} -\textstyle{\frac{1}{12}}\; H^2 -
     \textstyle{\frac{2}{3}} (\partial_\mu \Gamma_\nu -
\partial_\nu\Gamma_\mu)
B^{\mu\nu} - \bar{R}_{\mu\alpha\nu\beta}B^{\mu\nu}B^{\alpha\beta} +
{\cal O} (B^3)] \nonumber \\
& \equiv & \int{d^4x} \sqrt{-G} \bar{R} \, + \, I_M \, .
\label{Di}
\end{eqnarray}
In this formulation the matter stress tensor reads
\begin{eqnarray}
T_M^{\mu\nu} \; = \; 2 (-G)^{-1/2}
\textstyle{\frac{\delta}{\delta G_{\mu\nu}}} \;
I_M & = &
\left( \textstyle{\frac{1}{2}} H^{\mu\alpha\beta} H^\nu\,_{\alpha\beta} -
\textstyle{\frac{1}{12}} G^{\mu\nu} \; H^2 \right) \nonumber \\
&+& \left( \textstyle{\frac{4}{3}} B^{\mu\alpha} f^\nu\,_\alpha +
\textstyle{\frac{4}{3}} B^{\nu\alpha} f^\mu\,_\alpha -
\textstyle{\frac{2}{3}} \; G^{\mu\nu} \; f^{\alpha\beta} B_{\alpha\beta}
\right)  \nonumber \\
&+&
(3 \bar{R}^{(\mu}_{\;\;\gamma\alpha\beta}B^{\nu)\alpha}B^{\gamma\beta}
- G^{\mu\nu}\bar{R}_{\gamma\alpha\delta\beta}B^{\gamma\delta}B^{\alpha\beta})
\nonumber \\
&-& \bar{\nabla}_\beta\bar{\nabla}_\alpha (B^{\alpha(\mu}B^{\nu)\beta}) \, ,
\label{Dj}
\\
f_{\mu\nu} \; \equiv \; \partial_\mu \Gamma_\nu - \partial_\nu \Gamma_\mu \, .
\nonumber
\end{eqnarray}
Only the first term on the RHS leads to positive energy; its $00$ component
is (omitting metric factors) given by
$\sim 3 \; H^2_{0ij} + H^2_{ijk}$.  The remaining ones fail to do so,
even  after using the $B$ equations of motion (\ref{Df}) to
eliminate $f_{\mu\nu} = \frac{3}{4} \; (\bar{\nabla}^\lambda H_{\lambda\mu\nu}
 - 4\bar{R}_{\mu\alpha\nu\beta}B^{\alpha\beta})$.  Thus the proof of
positivity of energy for normal fields fails here.  Still
assuming that all fields fall off sufficiently fast at infinity, let us
consider the asymptotic energy flux associated with (\ref{Dj}).  At
infinity, the third term in (\ref{Dj}) vanishes, while the fourth one
is easily seen to be a simple superpotential ({\sl i.e.} of the
form $\partial_\alpha \partial_\beta K^{[\mu\alpha] [\nu\beta]}$)
making no net contribution to the energy-momentum flux.
On the other hand, the first term in (\ref{Dj}) gives a positive contribution
corresponding to the radiation of the one scalar mode contained in the
transverse projection of $B_{ij}$, while the second term gives
a non-positive contribution associated
with the interplay between the radiation of $B_{0i}$ and that of $\Gamma_i$.
When comparing our results with those of the related analysis in \cite{Kris},
we found sign mistakes between (2.21c) of \cite{Kris} and the
``NGT luminosity formula'' (3.16) of \cite{Kris}.
In fact, the signs of the NGT terms in (3.16) of \cite{Kris}
should be reversed to be made to agree with our (\ref{Dj}).  With
corrected signs the results (4.39) for the ``radiation
fields to lowest post-Newtonian order'' in \cite{Kris} imply that NGT
waves radiate a negative flux of dipole radiation.  However it must be
pointed out that such a calculation is even in itself undermined by
another dire physical consequence of NGT, namely the loss of any
acceptable falloff behaviour of $B_{\mu\nu}$ in the wave zone.

    To investigate the falloff behaviour, let us assume that the
curvature decays fast enough at infinity.  Then, (\ref{Dh}) yields
(in the Lorentz gauge $\bar{\nabla}^\mu\Gamma_\mu=0$) an inhomogeneous
wave equation for $\Gamma_\mu$ which implies that it has the
usual $1/r$ falloff at future null infinity.  However, inserting this
information in the r.h.s. of (\ref{Df}), one finds that $B_{\mu\nu}$
(and in particular its longitudinal part) fails to vanish at
future null infinity.  For example, we have checked explicitly
that in the usual matter-coupled version of NGT \cite{Moff},
considered in the post-Newtonian approximation \cite{Kris}, $\Gamma_\mu$
is predominantly radiated as a dipole, {\sl i.e.}
\begin{eqnarray}
\Gamma^0 & \sim & - \partial_i [ \partial_t^2 D^i(t-r)/r] \nonumber \\
\Gamma^i & \sim & \partial_t^3 D^i(t-r)/r \, ,
\label{Dka}
\end{eqnarray}
where $D^i$ is the dipole moment of the conserved particle number density
$S^0$ used as a macroscopic source in NGT.  This in turn generates
a nondecaying behaviour for $B_{\mu\nu}$, {\it e.g.}, for $r \rightarrow
\infty$
at constant $t-r$,
\begin{eqnarray}
B_{0i}({\bf r},t) & = & A(n_in_j-\delta_{ij})\partial_t^3D^j(t-r) +
{\cal O}(\textstyle{\frac{1}{r}}) \, , \nonumber \\
B_{ij}({\bf r},t) & = & A n_{[i} \partial_t^3 D_{j]} + {\cal O}(1/r) \, ,
\label{Dl}
\end{eqnarray}
where $n^i \equiv \frac{r^i}{r}$, and $A$ is a constant.   The problem
can only get worse at higher orders in $B_{\mu\nu}$, suggesting
that the fully nonlinear theory in $B$ does not make sense globally
either.\footnote{The error in the recent work \cite{MT} is readily recognized
by noticing that their antisymmetric field equation (4.4f) is an ordinary
differential equation for the function of one variable $\alpha(u)$ with
coefficients also depending on the other variables $r$ and $\theta$.
Therefore, in the general case of an arbitrary general relativistic
``news function'' $c(u,\theta)$, it admits only the trivial solution
$\alpha(u) = 0$.
[This conclusion already follows from a consideration of the leading terms
in the Bondi expansion of the antisymmetric field equation, as we have
checked explicitly.]}

  Let us note that the above inconsistencies are reminiscent of the
well-known problems with higher spin matter-gravity couplings
(see {\it e.g.}, \cite{AD}).
Like the latter they occur at the first nontrivial level beyond
flat space, and give rise to complicated causality structure.
Indeed, it is clear from the presence of the
$\bar{R}_{\mu\nu\alpha\beta}B^{\mu\alpha}B^{\nu\beta}$ term in the
action -- which couples
{\it second} derivatives of $G_{\mu\nu}$ with $B^2$ terms -- that the
characteristics of the theory will differ from the usual
(multiple) $G_{\mu\nu}$-light cone by $B^2$ terms.  In fact,
mathematical study of the exact characteristics of vacuum NGT
\cite{L55} \cite{MTV}, showed the existence
of {\it three} different characteristic cones along which coupled $G-B$ modes
propagate.

In summary, we conclude that since only $\bar{R}(G) + H^{2}$ is
consistent, and all natural two-derivative
`geometric' formulations typically have both unremovable naked
$B$-dependence and/or the wrong $(\bar{\nabla}B)^{2}$ structure,
Pauli's words about NGT,
``what God has put asunder, let no man set together"
apply quite aptly to all nonsymmetric gravity models without
cosmological terms.

\renewcommand{\theequation}{5.\arabic{equation}}
\setcounter{equation}{0}

\noindent{\bf 5. ~Finite--range models.}

We saw in the previous sections that ``geometric" nonsymmetric
theories
generically failed to obey consistency requirements unless the
antisymmetric field is given a finite range.\footnote{We noted that the
symmetric part of the field
need not itself acquire a
cosmological constant; in any case the latter is not relevant in the
search for distinctly non--Einsteinian effects.}
In this section we
consider these massive models with emphasis on their possible
observational consequences when coupled with
matter,
exploring ways in which the antisymmetric tensor  contributes to
the
effective gravitational interactions of matter at laboratory or
celestial scales.
Needless to say, we will keep to ghost--free models whose kinematic
terms are the gauge--invariant ${-\frac{1}{12}}\;
H^2_{\lambda\mu\nu}.$
To set some bounds on the range $\mu^{-1}$ (or mass $\mu$), which is
a free parameter, we recall
that its formal introduction was through a cosmological--type term
which
suggests an upper bound $\mu^{-1} \sim \Lambda^{-1/2} \sim H_0 \sim
10^{10}$ light years.
A lower bound will be
dictated
by the matter couplings chosen and the limitations due to technical
possibilities.

The basic question in studying alternative gravity models
is to identify plausible macroscopic sources of
$B_{\mu\nu}$. The first possibility which suggests itself is a direct
coupling of $B_{\mu\nu}$ to some kind of microscopic polarization
tensor, say $J^{\mu\nu} = \overline{\psi} \sigma^{\mu\nu} \psi$.
However, such a coupling yields negligible macroscopic forces between
ordinary (unpolarized) laboratory or celestial bodies, and
therefore does not constitute an interesting starting point
for our purposes.  A different
possibility is
suggested by the link with the geometrical $g_{\mu\nu} -
\Gamma^\lambda_{\ \mu\nu}$ models explored above. Indeed,
equation (\ref{Fe}) for instance shows that the models we want to explore
contain a propagating torsion
\begin{equation}
 T^\lambda_{\ \mu\nu} = \overline{\nabla}^\lambda B_{\mu\nu} -
 \frac{1}{2}\; H^{\lambda}_{\ \mu\nu} + {\cal O} (B^3) \ .
  \label{5.1}
\end{equation}
By analogy with the coupling of $G_{\mu\nu}$ to the stress--energy
tensor $T^{\mu\nu}$, it is natural to look for a 3--index macroscopic
source $S^{\mu\nu}_\lambda$ coupled to the torsion (\ref{5.1}) -- or,
by duality, a vector current coupled to
$\epsilon^{\mu\nu}_{\;\;\;\;\lambda\alpha} T^\lambda_{\,\mu\nu}$. Natural
candidates are Dirac currents,
$\overline\psi \gamma_\alpha \psi$
(the resulting coupling constant is dimensionless if $B_{\mu\nu}$,
like $G_{\mu\nu}$, is geometrically normalized). Let us denote a
general
combination of such fermion currents by $J^\mu$. The total
antisymmetry of the tensor density
$J^{\ast\lambda\mu\nu}\equiv \epsilon^{\lambda\mu\nu\alpha} J_\alpha$
implies that its coupling to the torsion (\ref{5.1}) involves only the field
strength $H_{\lambda\mu\nu}$. [Note that throughout the remainder of this
section the $\epsilon$-symbols will be simply a totally antisymmetric
collection of numerical constants, completely metric-independent.
Thus $\epsilon^{\lambda\mu\nu\alpha}$ transforms as a
contravariant tensor density, while
$\epsilon_{\lambda\mu\nu\alpha} \equiv
\epsilon^{\lambda\mu\nu\alpha} = \pm 1,0$
is {\it not} related by the metric.  When needed,
appropriate factors of $\sqrt{-G}$ will be inserted explicitly.]

The basic macroscopic coupling of the $B$
field that we shall consider is then
\begin{equation}
 S_J = \int d^4 x {\cal L}_J = \frac{-1}{6}\; f
   \int \;d^4x\,H_{\lambda\mu\nu} J^{*\lambda\mu\nu} =
   \frac{1}{2}\; f \int d^4 x\; \epsilon^{\mu\nu\alpha\beta}
   B_{\alpha\beta} \partial_\mu J_\nu \ ,  \label{5.2}
\end{equation}
where $f$ is dimensionless if
$B_{\mu\nu}$ is.  The Pauli
coupling (\ref{5.2}), being actually gauge--invariant for any $J_\nu$
(conserved or not), could also have been used to couple a massless (gauge)
$B$ field to matter.  However, it is easily seen that in that context
$B_{\mu\nu}$ is really a massless scalar, $\phi$,
coupled to $\partial_\mu J^\mu$,
which is merely another way of introducing an extra scalar
interaction.

 Having selected a basic macroscopic source for the $B$ field, we can
use the geometric motivation to add further phenomenological couplings
between matter and the $B$ field. In particular, using the general
field redefinition (\ref{Be}), we can define a class of metric
theories of gravity where matter also feels the influence of a
two--parameter ``physical" metric
\begin{equation}
\tilde g_{\mu\nu} = G_{\mu\nu} + a B_{\mu\alpha} B^{\ \alpha}_\nu +
   b B_{\alpha\beta} B^{\alpha\beta} G_{\mu\nu} + {\cal O}(B^4)\;. \label{5.3}
\end{equation}
Drawing on the example of the Einstein--Cartan--type theories, with
independent torsion, propagating or not (see {\it e.g.},
\cite{HNH80}), we can
also consider more general theories where matter is coupled to
the above $\tilde g_{\mu\nu}$, as well as to some independent
connection
with torsion, say
\begin{equation}
\hat{\Gamma}^\lambda_{\mu\nu} = \{{}^\lambda_{\mu\nu}\}_{\tilde g} +
    c \widetilde\nabla^\lambda B_{\mu\nu} + d \tilde g^{\lambda\alpha}
H_{\alpha\mu\nu} +  \cdots \; , \label{5.4}
\end{equation}
where the first term on the right--hand side denotes the Christoffel
connection of $\tilde g_{\mu\nu}$ and where $c$ and $d$ are two additional
parameters.  Further terms could represent similar torsion terms
built with the dual of $B_{\mu\nu}$, and/or higher order terms.

 If all matter were minimally coupled to $\tilde g_{\mu\nu}$, then the
equivalence principle would still hold (apart from the current--current
finite range interaction entailed by the direct coupling (\ref{5.2})).
However, use of the full connection (\ref{5.4}) (required for fermions)
or allowing matter couplings other than just to $\tilde g_{\mu\nu}$
generically violates the equivalence principle. A convenient
parametrization of equivalence violation is to allow different types of
matter to couple to different ``metrics" $\tilde g^A_{\mu\nu}$ and
affinities $\hat{\Gamma}^{A\lambda}_{\mu\nu}$ labelled by different parameters
$(a_A, b_A)$ in (\ref{5.3}) and $(c_A, d_A)$ in (\ref{5.4}).

 A stronger, dilaton--inspired way of violating the equivalence
principle would be to have the material Lagrangians multiplied by
factors of the form $(1+eB^{\alpha\beta} B_{\alpha\beta})$. Finally,
recent studies in NGT \cite{W89}, \cite{MPM89}, \cite{GHMP91a},
also suggest the
possibility of violating the equivalence principle by adding non--metric
couplings between $B_{\mu\nu}$ and the electromagnetic field, say
\begin{equation}
  {\cal L}'_{BF} = \alpha (B_{\mu\nu} F^{\mu\nu})^2 \; . \label{5.5}
\end{equation}
[The other possible quadratic couplings $B^{\mu\alpha} B^\nu_{\ \alpha}
F_{\mu\beta}
F_\nu^{\ \beta}$ and $B^{\alpha\beta}B_{\alpha\beta}F_{\mu\nu}F^{\mu\nu}$
are already covered
by the parameters $a$ and $e$ introduced above, and the terms quadratic
in the dual $F^\ast_{\mu\nu}$, {\it e.g.}, $(B_{\mu\nu} F^{\ast\mu\nu})^2$,
are equivalent to these; a linear
mixing $B_{\mu\nu}F^{\mu\nu}$ would lead to unacceptable
asymptotic behavior].
Note that even if there is only one metric $\tilde g_{\mu\nu}$ which
couples universally to matter, then (\ref{5.5}) violates not only
equivalence but also local Lorentz invariance and local isotropy of
space.  Indeed, by Eq (\ref{5.2}), fermionic currents outside
a local laboratory will generate an external, macroscopic, $B_{\mu\nu}$ field
that will introduce, via (\ref{5.5}) preferred spacetime directions in all
physical effects involving electromagnetism.

We propose, then, a general multi--parameter gravity model
defined by a Lagrangian of the type
\begin{equation}
{\cal L} = {\cal L}_G + {\cal L}_B + {\cal L}_J +
  \sum_A (1+e_A B^2) {\cal L}_A
\left[ \psi_A, \tilde g^A_{\mu\nu}, \hat\Gamma^{A\lambda}_{\mu\nu} \right]
  + {\cal L}'_{BF} \; , \label{5.6}
\end{equation}
where
\begin{equation}
{\cal L}_G = \frac{\sqrt{-G}}{4\kappa^2}\;  \overline{R} (G) \; , \label{5.7}
\end{equation}
is the Einstein Lagrangian ($\kappa^2 \equiv 4 \pi G_{\rm Newton}$),
\begin{equation}
 {\cal L}_B = - \frac{\sqrt{-G}}{\kappa^2}\;
  \left[ \frac{1}{12}\; H^2 + \frac{1}{4}\; \mu^2 B^2 \right] \; ,
   \label{5.8}
\end{equation}
is the massive ghost--free $B$ Lagrangian (with $B$ dimensionless in
these gravitational units), and
${\cal L}_A$ denotes the Lagrangian describing some matter field
$\psi_A$ minimally coupled to a metric and a connection, {\it e.g.},
\begin{equation}
 {\cal L}_{\rm Dirac} = \sqrt{-G}\; \overline{\psi} \gamma^a
 \left( \tilde e^\mu_a \partial_\mu + \textstyle{\frac{1}{2}}\;
  \hat\Gamma^b_{ca} \sigma^c_b \right) \psi \; . \label{5.9}
\end{equation}
Here $\tilde e^\mu_a$ is the vierbein and $\hat\Gamma^{b}_{ac}$ a general
connection (\ref{5.4}); latin letters denote local indices. The
other terms in (\ref{5.6}) denote the coupling terms already
introduced, (\ref{5.2}) and (\ref{5.5}). The only dimensionful
``gravitational" parameters entering the general class of Lagrangians
(\ref{5.6}) are the bare gravitational coupling constant $\kappa$ and the
mass $\mu$ of $B$ (with dimensions of inverse length
in ordinary units). The dimensionless parameters are the (possibly
matter--dependent) numbers $(a_A,\cdots, e_A),$ $\alpha$, together with
the
current coupling constant $f$ and various mixing angles defining the
given current. Such angles have been discussed in the ``fifth force"
literature, where it was noted that the current $J^\mu$
could contain several free weight parameters ({\it e.g.}
distinguishing the lepton numbers belonging to different families).
In the case of coupling to ordinary macroscopic matter, it is enough
to introduce one mixing angle \cite{DR86}, say $\theta_5$, such
that the
total ``charge" is
\begin{equation}
 q_5 \equiv \int \sqrt{-G}\; d^3x J^0 = B \cos (\theta_5)
 + L \sin (\theta_5) \; , \label{5.10}
\end{equation}
where $B=N+Z$ is the baryon number and $L=Z$ the total lepton
number.
[When exploring cosmological aspects of the present class of theories,
one also has the possibility of introducing a different coupling
to fermionic dark matter.]

We certainly do not claim that such a general model recommends itself
by its elegance, but we propose it as
a replacement for NGT to provide a
foil for general relativity,
useful for suggesting some intriguing possibilities in experimental
gravitation.
In addition,
our general approach, motivated as it is by string theory,
non--symmetric geometric models and Einstein--Cartan--type theories,
is likely to encompass more possibilities.

In order to outline the main types of observational consequences of our
general model, let us study the $B$ field equations. If $K_{\mu\nu}$
denotes the effective source for $B_{\mu\nu}$ (including nonlinearities),
{\it i.e.},
\begin{equation}
 K^{\mu\nu} = \frac{2}{\sqrt{-G}}\; \frac{\delta}{\delta B_{\mu\nu}}\;
 \left[ {\cal L}_J + \sum_A (1+e_AB^2) {\cal L}_A +
 {\cal L}'_{BF} \right]\; , \label{5.11}
\end{equation}
we can formally write the $B$ field equations as
\begin{equation}
 - \overline\nabla^\lambda H_{\lambda\mu\nu} + \mu^2 B_{\mu\nu} =
  \kappa^2 K_{\mu\nu} \; . \label{5.12}
\end{equation}
The identity
$\overline\nabla^\lambda
\overline\nabla^\mu H_{\lambda\mu\nu} \equiv 0$ implies
\begin{equation}
 \overline\nabla^\nu B_{\mu\nu} = \frac{\kappa^2}{\mu^2}\;
 \overline\nabla^\nu K_{\mu\nu} \; , \label{5.13}
\end{equation}
and hence we obtain the following inhomogeneous Klein--Gordon
equation\footnote{Let us note in passing the singularity of the massless
limit as made evident by the $1/\mu^2$ factors in the right--hand
sides of (\ref{5.13}) and (\ref{5.14}). However, this singular behavior
concerns mainly nonlinear ${\cal O}(\kappa^4)$ terms in $B$ because, as
is
clear from (\ref{5.2}), the leading source term, $\sqrt{-G} K^{\mu\nu} =
f\,\partial_{\!\lambda} J^{\ast\lambda\mu\nu} + {\cal O}(\kappa^2 f)$,
is conserved:
$\overline\nabla_{\!\nu} K^{\mu\nu} = 0 + {\cal O}(\kappa^2 f)$.}
for $B_{\mu\nu}$
\begin{equation}
  (\bar{\Delta} B)_{\mu\nu} + \mu^2 B_{\mu\nu} = \kappa^2
\left( K_{\mu\nu} + \;\frac{1}{\mu^2}\; \partial_{[\mu}
 \overline\nabla^\alpha K_{\nu ]\alpha} \right)\; , \label{5.14}
\end{equation}
where $\bar{\Delta}$ denotes the Hodge--de~Rham Laplacian in the metric
$G_{\mu\nu}$, differing from the ordinary Laplacian by curvature terms:
\begin{equation}
  (\bar{\Delta} B)_{\mu\nu} \equiv - G^{\alpha\beta}\;
\overline\nabla_{\!\alpha}\;    \overline\nabla_{\!\beta}
   B_{\mu\nu}\;+\; (\bar{R}B)_{\mu\nu} \, .    \label{5.15}
\end{equation}

If $K_{\mu\nu}$ were a linear ($B$--independent) source, the interaction
mediated by $B_{\mu\nu}$ would be equivalent to that due to a massive
vectorial
(Proca) field. This equivalence is most easily seen in first order
formalism. Indeed, let us start from the first order form of the massive
$B$ theory with both potential and field--strength (Pauli--type)
coupling to external sources,
\begin{eqnarray}
 {\cal L}(B,H) = - \frac{\sqrt{-G}}{\kappa^2}
&& \left[ - \frac{1}{12}\;H^2 + \frac{1}{4}\;\mu^2 B^2 +\;
  \frac{1}{6} H^{\lambda\mu\nu} \partial_{[\lambda} B_{\mu\nu ]} \right]
\nonumber \\
&& + \frac{1}{2}\;\sqrt{-G}\; B_{\mu\nu} K^{\mu\nu}
   + \frac{1}{6}\;\sqrt{-G}\; H_{\lambda\mu\nu} L^{\lambda\mu\nu} \; .
\label{5.16}
\end{eqnarray}

Defining\footnote{Note again the singularity of the massless limit,
which we know turns the $B$ field into a massless scalar.}
$F_{\mu\nu} =  \sqrt{-G}\;\frac{\mu}{2\kappa} \epsilon_{\mu\nu\alpha\beta}
B^{\alpha\beta}$,
{}~$A_\mu = - \frac{\sqrt{-G}}{6\mu\kappa} \epsilon_{\mu\alpha\beta\gamma}
H^{\alpha\beta\gamma}$, one
obtains
\begin{eqnarray}
 {\cal L}(A,F) = \sqrt{-G}\;
&& \left[ \frac{1}{4}\;F^2 - \frac{1}{2}\;\mu^2 A^2 - \frac{1}{2}\;
   F^{\mu\nu} \partial_{[\mu} A_{\nu ]} \right] \nonumber \\
&& + \sqrt{-G}\; A_\mu J^\mu_{\rm Proca}
   + \frac{1}{2}\;\sqrt{-G}\; F_{\mu\nu} M^{\mu\nu}_{\rm Proca}\; ,
   \label{5.17}
\end{eqnarray}
which is the first order form of a Proca field coupled to
\begin{equation}
 J^\mu_{\rm Proca} =\; - \frac{\kappa\mu}{6}\;
\frac{\epsilon^{\mu\alpha\beta\gamma}}{\sqrt{-G}} \; L_{\alpha\beta\gamma} \, ,
 \,\,\, M^{\mu\nu}_{\rm Proca} =\;\frac{\kappa}{2\mu}\;
\frac{\epsilon^{\mu\nu\alpha\beta}}{\sqrt{-G}} \; K_{\alpha\beta} \, .
 \label{5.18}
\end{equation}

  Equation~(\ref{5.14}) can formally be solved by successive iterations
in powers of $\kappa^2$,
\begin{equation}
 B_{\mu\nu} = \kappa^2 B^{(1)}_{\mu\nu} + \kappa^4 B^{(2)}_{\mu\nu} +\cdots ,
 \,\,\,\,  G_{\mu\nu} = G^{(0)}_{\mu\nu} + \kappa^2 G^{(1)}_{\mu\nu} +
\cdots \; ,
 \label{5.19}
\end{equation}
where, for simplicity, we shall consider a flat background metric,
$G^{(0)}_{\mu\nu} = \eta_{\mu\nu}$. Inserting the ${\cal O}(\kappa^0)$
source read off  from (\ref{5.2}),
\begin{equation}
 K_{\mu\nu} =\; - f\; \epsilon_{\mu\nu\alpha\beta} \partial^\alpha J^\beta+\;
  {\cal O} (\kappa^2 f) \; , \label{5.20}
\end{equation}
one gets
\begin{equation}
  B_{\mu\nu} = \kappa^2 f (\Box -\mu^2)^{-1}
  [ \epsilon_{\mu\nu\alpha\beta} \partial^\alpha J^\beta] +
{\cal O} (\kappa^4)\; . \label{5.21}
\end{equation}
Moreover, our previous results (\ref{5.15})--(\ref{5.17}) [remembering
the Pauli form (\ref{5.2}) of the lowest order matter coupling]
show that the lowest order matter--matter coupling mediated by the
$B$ field (\ref{5.21}) is equivalent (apart from contact terms) to a
vectorial ``fifth force" interaction coupled to the fermion current
$J^\mu$ [with total charge (\ref{5.10})] with dimensionless coupling
constant
\begin{equation}
 g_5 = \kappa \mu f \; .  \label{5.22}
\end{equation}
Note the proportionality of the strength of the effective fifth force
coupling to the inverse range of the interaction. On the other hand, recent
experimental work on possible fifth forces
has put bounds on $g_5$ that, though becoming more stringent as the
range $\mu^{-1}$ increases, stay finite as $\mu$ vanishes. To quote some
numbers, the dimensionless quantity $\alpha_5 = g^2_5 / (\kappa^2 m^2_N)$
(where $m_N$ denotes, say, one atomic mass unit) was found to be bounded
by $\hbox{\raise.5ex\hbox{$<$} \kern-1.1em\lower.5ex\hbox{$\sim$}} 10^{-3}$
when $\mu^{-1} = 1$~m, $\hbox{\raise.5ex\hbox{$<$}
  \kern-1.1em\lower.5ex\hbox{$\sim$}} 10^{-5}$ when $\mu^{-1}
= 1$~km, and $\hbox{\raise.5ex\hbox{$<$}
  \kern-1.1em\lower.5ex\hbox{$\sim$}} 10^{-8}$
for $\mu^{-1} > 10^4$~km (assuming a coupling
to baryon number, i.e. $\theta_5 = 0$ in (\ref{5.2}); see \cite{A90}
for precise numbers, the dependence on the mixing angle and further
references). The important points for our present phenomenological
purpose are that the strength $f$ of our basic coupling is unbounded
as the range increases, and that the magnitude (\ref{5.21}) of the $B$
field itself is primarily proportional to $f$ and independent
of the range $\mu^{-1}$ (until the nonlinear terms $\propto \kappa^2
\mu^{-2}$ come into play). This therefore opens the possibility
of having a $B$ field of ``gravitational" strength, contributing
significant new macroscopic forces, while still keeping compatibility
with
the existing stringent bounds on possible composition--dependent effects
in Newtonian
gravity.\footnote{Note that, in view of our above proof of
equivalence of linear $B$--couplings to vectorial ones, this possibility
exists only if there are explicit $B$--dependent terms in the ${\cal
L}_A$--type Lagrangians (i.e. some parameters among
$a_A$,\dots,$e_A$,
$\alpha$ must be non zero).}

 To discuss the leading observational consequences of our general
framework, it is convenient to replace the fundamental dimensionless
coupling constant $f$ by the auxiliary quantity
\begin{equation}
 \lambda \equiv \frac{f}{m_N}\;\approx\; (2\, f) \times 10^{-14}  \, {\rm cm}
   \label{5.23}
\end{equation}
which has the dimension of length and which couples $B$ to $m_N
J^\mu$,
{\it i.e.}, roughly speaking, to the baryon rest--mass density. Then we
introduce, as leading ``short--range gravitational potential", the
quantity
\begin{equation}
  V \equiv \; \frac{\kappa^2 m_N J^0}{\mu^2 - \Box}  \label{5.24}
\end{equation}
(which is approximately equal to the Newtonian potential
$U \equiv -\kappa^2 \Delta^{-1} T^{00}$ when the range is large).
In units where only the velocity of light is set equal to one, the
potential $V$ is, like $U$, dimensionless. Then the leading components
of $B_{\mu\nu}$ (for slowly moving sources $J^i/J^0 \sim v \ll 1$) are
the ``magnetic" ones,
\begin{equation}
  B^\ast_{0i} = \; \frac{1}{2}\;\epsilon_{ijk} B_{jk} = \; \lambda \,
\partial_i V\; ,  \label{5.25}
\end{equation}
the electric components $B_{0i} =\,\frac{1}{2}\;\epsilon_{ijk} B^\ast_{jk}$
being $v$ times smaller.

 We can reexpress the auxiliary coupling length $\lambda$ in terms of
experimentally constrained fifth force quantities as
\begin{equation}
  \lambda = \sqrt{\alpha_5}\; \mu^{-1} = \; \sqrt{\alpha_5} \times \
  (\hbox{range of}\ B).    \label{5.26}
\end{equation}
This shows that $\lambda$ can take any macroscopic value, as long as the
range of $B$ is large enough (the corresponding value of $f$ is then
large, {\it e.g.}, $f\sim 10^{19} \sim m_P/m_N$ for $\lambda \sim 1\; {\rm
km}$, where $m_P$ denotes the Planck mass).

At order $\kappa^4$, the $B$ field has two kinds of metric--gravitational
effects. On the one hand the ${\cal L}_B$--part of the action
contributes ${\cal O}(B^2)$ source terms in the Einstein equations for
$G_{\mu\nu}$, and on the other hand the $B^2$ terms in the definition
(\ref{5.3}) contribute directly to the physical metric in which each
type of matter ``falls''. Both types of terms will contribute terms of the
form
\begin{equation}
  -\;\frac{1}{2}\;{\rm ln} (-\tilde g_{00})\;=\;
   U+k (\lambda \partial_i V)^2 + \cdots\;,  \label{5.27}
\end{equation}
for some numerical constant $k$, to the logarithm  of
the time--time component of the metric (which plays the role of a
``quasi--Newtonian" potential). [We have omitted $\mu$--dependent
terms on the right that we do not attempt to discuss here]. Hence, we
see that,
when $\mu^{-1}$ is greater than the characteristic distances of the
problem under consideration, there will be a ``van der Waals'' higher
power type
contribution $\sim \kappa^4 \lambda^2 M^2/r^4$ to the effective gravitational
potential (this arises separately from the $\sim \kappa^2\lambda^2 \mu^2 M/r$
composition--dependent fifth force potential).

 If we consider the minimal models with a universal metric coupling
$(a_A=a,$ $b_A=b$ for all $A$'s, $c_A=d_A=e_A=\alpha=0)$ the main new
effects carried by the $B$ field [apart from the fifth force one]
will be associated  with the $(\lambda{\bf\nabla}V)^2$ potential
of~(\ref{5.27})
[as stated above, this requires nonzero explicit $B^2$ terms in the
physical
metric; $a^2+b^2\not= 0$]. As $\lambda$ is a universal length, these effects
should be strongest in the smallest and most strongly self--gravitating
$(V \hbox{\raise.5ex\hbox{$<$} \kern-1.1em\lower.5ex\hbox{$\sim$}} U\sim 1)$
objects, namely neutron stars. Such models could
therefore offer interesting foils for the strong--field regime of
general relativity. A detailed study is needed to see if they could
be tested by means of binary pulsar data, following the methodology
of \cite{DT92}.
If $\lambda$ is large enough these terms could be tested in the solar system
as $(\lambda/r)^2$ fractional deviations of the post--Newtonian effects. In
this, and the following, discussion we assume a range larger
than the characteristic distances of the system considered. We are
aware
that this might be difficult to achieve beyond certain limits because of
the formal $\mu^{-2}$ singularity apparent in (\ref{5.12}), (\ref{5.13}).
A closer study of what happens when $\mu$ becomes very small is
needed.

 If we turn now to the most general models of the type~(\ref{5.6})
they will predict, beyond the effects already discussed,
\noindent
(i) violations of the weak equivalence principle, when the
$a_A$'s and $b_A$'s are not the same for all material fields or
interactions, or when $e_A \not= 0$. These violations will be
proportional
to the $(\lambda {\bf \nabla} V)^2$ type of terms.
\noindent
(ii) violations of local Lorentz invariance and existence of
peculiar
electromagnetic phenomena when, $\alpha \not= 0$, as recently pointed out
in \cite{GHMP91a}, \cite{GHMP91bc}.
\noindent
(iii) ``monopole--dipole" coupling of the (quantum) spin of
elementary particles to the macroscopic $B$ field generated by, say, the
baryons in the Earth. The latter coupling is akin to those proposed
in \cite{LOHDMWG}. In our model this type of coupling comes from the possible
additional torsion terms in~(\ref{5.4}). Indeed, a Dirac particle couples to
torsion via
\begin{equation}
  \frac{1}{2}\ \sqrt{-G} \ \overline\psi \ \gamma^\mu T^\alpha_{\beta \mu}
  \sigma_{\alpha .}^{\ \beta} \psi \; . \label{5.28}
\end{equation}
One can see that the terms written in~(\ref{5.4}) give an interaction
of order $v\lambda \partial^2 V {\bf \sigma}$ where ${\bf \sigma}$ is the
spin and $v$
a small velocity.  Had one introduced terms $c_\ast \widetilde\nabla^\lambda
B^\ast_{\mu\nu} + d_\ast\;\tilde g^{\lambda\alpha} \partial_{[\alpha}
B^\ast_{\mu\nu]}$
on the
right-hand side of (\ref{5.4}), the coupling would have been of the full order
$\lambda \partial^2 V {\bf \sigma}$ without velocity factors.
As was briefly mentioned at the beginning of this section, one could
also think of adding non--derivative couplings of $B_{\mu\nu}$ to some
microscopic polarization tensor, {\it i.e.} $ \sim B_{\mu\nu}(\overline\psi
\sigma^{\mu\nu} \psi)$. The latter coupling would contribute
interactions
proportional to ${\bf \nabla} V {\bf \sigma}$.

In summary, our new gravity models incorporating a finite range antisymmetric
tensor field are consistent, and they provide
a vast reservoir of interesting phenomenological
possibilities in experimental gravitation.

Acknowledgements: T.D. thanks G.W. Gibbons for fruitful discussions in the
early stages of this work.  We are grateful to A. Lichnerowicz
for an informative conversation on earlier literature.
The work of S.D. and J.M. was supported in part by NSF grant PHY88-04561;
S.D thanks IHES for hospitality.

\newpage
\renewcommand{\theequation}{A.\arabic{equation}}
\setcounter{equation}{0}

\noindent {\bf Appendix A.  ~The $B_{\mu\nu}$ expansion.}

 From the general form of $g_{\mu\nu}$,
\begin{equation}%
g_{\mu\nu} = G_{\mu\nu} + B_{\mu\nu} + \alpha B_{\mu\alpha}
B^\alpha_{\;\;\nu} +
\beta\: B^{2} G_{\mu\nu} + {\cal{O}}(B^3 )
\label{Fa}
\end{equation}
we easily determine its inverse and determinant,
\begin{equation}
g^{\mu\nu} = G^{\mu\nu} + B^{\mu\nu} + (1-\alpha) B^{\mu\alpha}
B_\alpha^{\;\;\nu} - \beta\:
B^{2} G^{\mu\nu} + {\cal{O}}(B^3 ) \; .
\label{Fb}
\end{equation}
\begin{equation}
\sqrt{-g} = \sqrt{-G}
\left[ 1 + \textstyle{\frac{1}{2}} \left(\textstyle{\frac{1}{2}} - \alpha +
\beta D \right) B^{2} \right] + {\cal{O}}(B^4) \; .
\label{Fc}
\end{equation}
Here we are using the notation $B^{\mu}_{\;\;\nu} \equiv
G^{\mu\alpha}B_{\alpha\nu} =
- B_\nu^{\;\;\mu}$, and $B^2 \equiv B^{\mu\nu}B_{\mu\nu}$.
Consider the defining relation for the affinity as a function of the
metric,
\begin{equation}
\partial_\lambda g_{\mu\nu} - \Gamma^\alpha_{\mu\lambda} \: g_{\alpha\nu} -
\Gamma^\alpha_{\lambda\nu} \: g_{\mu\alpha} = 0 \; .
\label{Fd}
\end{equation}
On introducing the expansion
$\Gamma^\lambda_{\mu\nu} = \sum_{n\geq 0} (S^{(n)\lambda}_{\;\;\mu\nu} +
A^{(n)\lambda}_{\;\;\mu\nu})$, (where the superscript $(n)$ denotes the
order in
$B_{\mu\nu}$, while $S$ and $A$ denote symmetric and
antisymmetric
parts
respectively) into (\ref{Fd}), we easily find
$A^{(0)\lambda}_{\;\;\mu\nu} =0$ and $S^{(0)\lambda}_{\;\;\mu\nu} =
\{ \textstyle{\stackrel{\scriptstyle\lambda}
{\scriptstyle\mu\nu}} \} (G)$, the Christoffel symbol
with
respect to the ``background" $G_{\mu\nu}$.  Matching higher orders,
we
have $S^{(1)\lambda}_{\;\;\mu\nu} = 0$ and
\begin{equation}
A^{(1)\lambda}_{\;\;\mu\nu} = \textstyle{\frac{1}{2}}
(\bar{\nabla}^\lambda B_{\mu\nu} -
\bar{\nabla}_\mu B_\nu^{\;\;\lambda} - \bar{\nabla}_\nu B^\lambda_{\;\;\mu} )
\label{Fe}
\end{equation}
where, as always, the bar denotes operations with respect to the
background
$G_{\mu\nu}$.  The higher powers are defined by
\begin{eqnarray}
S^{(2n)\alpha}_{\;\;\mu\lambda} \: G_{\alpha\nu} +
S^{(2n)\alpha}_{\;\;\lambda\nu} \: G_{\mu\alpha} & = &
-A^{(2n-1)\alpha}_{\;\;\mu\lambda} \: B_{\alpha\nu} -
A^{(2n-1)\alpha}_{\;\;\lambda\nu} \: B_{\mu\alpha}  \nonumber \\
A^{(2n+1)\alpha}_{\;\;\mu\lambda} \: G_{\alpha\nu} +
A^{(2n+1)\alpha}_{\;\;\lambda\nu} \: G_{\mu\alpha} & = &
-S^{(2n)\alpha}_{\;\;\mu\lambda} \: B_{\alpha\nu} -
S^{(2n)\alpha}_{\;\;\lambda\nu} \: B_{\mu\alpha} \; .
\label{Ff}
\end{eqnarray}
For simplicity we have written the equations (\ref{Ff}) as if
$g^{(n)}_{\mu\nu} = 0, \; n \geq 2$ in (\ref{Be}); the modification for
non-zero field redefinition is trivially accomplished, and does
not alter (\ref{Fe}).  Clearly (\ref{Ff}) may be solved
iteratively, to obtain the expansion of $\Gamma^{\lambda}_{\mu\nu}$ in powers
of $B_{\mu\nu}$.

The more general equation (\ref{Bo}) for the affinity
leads to a similar expansion.  Using it, we obtain to lowest order
\begin{equation}
\left( 1 + \frac{a+b+c+d}{2} \right) \;
A^{(0)\alpha}_{\;\;\mu [\lambda}  G_{\nu ]\alpha} = 0 \; .
\label{Fg}
\end{equation}
Thus generically we find
$A^{(0)\lambda}_{\;\;\mu\nu} = 0$ and, going on as
before, obtain an iterative solution.
The exceptional case where the affinity cannot be determined is exactly
that listed in the text, $ a+b+c+d = -2$.

With the above results we may easily write the expansion of the
geometric quantities defined in the text.  In particular, we find
\begin{eqnarray}
R_{\mu\nu}(g) & = & \bar{R}_{\mu\nu}(G) + \textstyle{\frac{1}{2}}
\bar{\nabla}^\alpha\bar{\nabla}_\alpha B_{\mu\nu} -
\textstyle{\frac{1}{2}} \bar{\nabla}^\alpha\bar{\nabla}_{[\mu}B_{\nu ]\alpha} -
\bar{\nabla}_\nu\bar{\nabla}^\alpha B_{\mu\alpha} +
{\cal{O}}(B^{2})\nonumber
\\
T^\lambda_{\mu\nu} & = & A^{(1)\lambda}_{\;\;\mu\nu} + {\cal{O}}(B^3)
\nonumber \\
\Gamma_\mu & = & \bar{\nabla}^\alpha B_{\mu\alpha} + {\cal O}(B^3) \nonumber
\\
\sqrt{-g} \:g^{\mu\nu} R_{\mu\nu}(g) & = &
\sqrt{-G} \: \bar{R} (G) - \textstyle{\frac{1}{12}} \:
\sqrt{-G} \: H^{\mu\nu\lambda}H_{\mu\nu\lambda} +
\textstyle{\frac{1}{2}} (\textstyle{\frac{1}{2}} - \alpha + (D-2)\beta)
\sqrt{-G} \: \bar{R}(G)B^{2} \nonumber \\
&  & - ~\alpha \sqrt{-G} \:
\bar{R}_{\mu\nu} (G)B^{\mu\alpha}B_\alpha^{\;\;\nu} -
\sqrt{-G} \: \bar{R}_{\mu\nu\alpha\beta} \: B^{\mu\alpha}B^{\nu\beta}
\nonumber \\
&+& \makebox{total derivative} + {\cal O}(B^4)
\nonumber \\
\sqrt{-g} \:g^{\mu\nu} \Gamma_\mu\Gamma_\nu & = & \sqrt{-G} \:G^{\mu\nu}
\bar{\nabla}^\alpha B_{\mu\alpha}\: \bar{\nabla}^\beta\: B_{\nu\beta}
+ {\cal{O}}(B^4) \nonumber \\
\sqrt{-g} \: g^{\mu\nu} \: T^\alpha_{\mu\lambda}T^\lambda_{\alpha\nu} & = &
-\textstyle{\frac{1}{12}} \sqrt{-G} \: \bar{H}^{\mu\nu\lambda}
\bar{H}_{\mu\nu\lambda} - \sqrt{-G} G_{\mu\nu}\: \bar{\nabla}_\beta\:
B^{\mu\alpha} \bar{\nabla}_\alpha \: B^{\nu\beta} + {\cal O}(B^4) \; .
\label{Fh}
\end{eqnarray}
We also use implicitly the fact, which follows from the
``$g^{\mu\nu}$--trace'' of (\ref{Fd}), that
\begin{equation}
\Gamma^\alpha_{(\lambda\alpha )} (g) = \partial_\lambda \: \ln \: (- g) \; ,
\label{Fi}
\end{equation}
and so
\begin{equation}
P_{\mu\nu} (\Gamma (g) ) = - \partial_{[\mu}\Gamma_{\nu ]} (g) \; .
\label{Fj}
\end{equation}
It is then easy to show that in second order, up to a total
derivative,
\begin{equation}
\sqrt{-g} \: g^{\mu\nu} \: P_{\mu\nu} = -2 \sqrt{-g} \: g^{\mu\nu}
\:
\Gamma_\mu\Gamma_\nu \; .
\label{Fk}
\end{equation}

\newpage

\renewcommand{\theequation}{B.\arabic{equation}}
\setcounter{equation}{0}

\noindent {\bf Appendix B.  ~First and Second Order Forms of NGT.}

The first order NGT action may be taken as
\begin{equation}
{\cal L}^{(1)} = \sqrt{-g} \: g^{\mu\nu} \: R_{\mu\nu} (\Gamma ) \; .
\label{Ga}
\end{equation}
We saw in Section 2 that its equations of motion are equivalent to those
of the second order action
\begin{equation}
{\cal L}^{(2)} = \sqrt{-g} \: g^{\mu\nu} \: R_{\mu\nu} (g) -
b_\nu \: \partial_\mu \: (\sqrt{-g} g^{[\mu\nu ]}) \; .
\label{Gb}
\end{equation}
The difficulties of this model at quadratic order and beyond are chronicled
in
detail in Section 4.  Our aim here is rather to clearly demonstrate the
equivalence of the above two formulations, since doubts have been
raised on this question \cite{Moff}.
It will suffice to do so for the linearized theory about a flat
background since this exhibits the kinematics of degrees of freedom
which
is at issue.  We begin with the second order action, which we expand in
powers of
$g_{\mu\nu} - \eta_{\mu\nu} = s_{\mu\nu} + B_{\mu\nu}$.  Using
Appendix A we find
\begin{equation}
{\cal L} = {\cal L}^{(2)}_E(s) + {\cal L}^{(2)}_B(B) + {\cal{O}}((B,s)^3)
\label{Gc}
\end{equation}
where ${\cal L}_E$ is the linearized Einstein action and
\begin{equation}
{\cal L}^{(2)}_B = - \textstyle{\frac{1}{12}} \: H^{2}_{\mu\nu\lambda} -
2 b_\nu\:\partial_\mu \:B^{\mu\nu} \; .
\label{Gd}
\end{equation}
This may be recognized as a gauge fixed form of the $H^2$ action akin
to
the Nakanishi--Lautrup \cite{NL} gauge fixing in
electrodynamics; its degrees of freedom are
well understood.  For completeness, we summarize the analysis of the
equations of motion from (\ref{Gd}), which are equivalent to
\begin{eqnarray}
\Box B_{\mu\nu} + 2 \partial_{[\mu} b_{\nu]} \; & = & \; 0 \label{Ge} \\
\partial^{\alpha} B_{\mu\alpha} \; & = & \; 0 \; . \label{Gf}
\end{eqnarray}

The divergence of (\ref{Ge}) states that $b_\mu$ obeys Maxwell's equations
\begin{equation}
\partial^\mu \: \partial_{[\mu}b_{\nu]} \; = \; 0 \; .
\label{Gg}
\end{equation}
Suppose now that we are given the $2 \times (D-2)$ Cauchy data
(corresponding to $(D-2)$ dynamical degrees of freedom) required to
specify (modulo its gauge invariance) $b_\mu$ from (\ref{Gg});
then (\ref{Ge}) implies
\begin{equation}
\Box \partial^\mu B_{\mu\nu} = 0 \; ,
\label{Gh}
\end{equation}
and so the constraint (\ref{Gf}) can be incorporated as a set of relations
among
the Cauchy data of (\ref{Ge}).  That is, requiring
$\partial^\alpha B_{\mu\alpha}$ and its
first time derivative to vanish at $t=0$ ensures that it obeys (\ref{Gf}) for
all time.  In turn, these relations determine the Cauchy data for $B_{0i}$
according to
\begin{eqnarray}
(\partial_0 B_{0i} & = & \partial^j B_{ji} ) |_{t=0} \nonumber \\
(\Delta B_{i0} & = & \partial^j \partial_0 B_{ij} - 2 \partial_i b_0 +
2 \partial_0 b_i ) |_{t=0}
\; ,
\label{Gi}
\end{eqnarray}
(from which $\partial^i B_{0i} = 0$ follows).
Hence the independent Cauchy data are the $2(D-2)$ for $b_\mu$
together with the
$(D-1)(D-2)$ $\{B_{ij}, \dot{B}_{ij}\}$.
Thus the total number of degrees of freedom is
$\textstyle{\frac{1}{2}} (D+1)(D-2)$,
namely 5 in $D=4$. [This is in agreement with an early estimate by Einstein,
third reference of \cite{EnO}.]  The apparent
paradox -- that this cannot be the correct count for a gauge-fixed theory
known to have but one degree of freedom in
$D=4$ -- is cleared up by the following
remarks.  First, in the gauge fixing, $b_\mu$ may be eliminated initially
(and hence consistently for all time).  The other is that (\ref{Gf}) has a
residual gauge invariance under
$\delta B_{\mu\nu} = \partial_{[\mu}\epsilon_{\nu]}$,
for $\epsilon_\mu$ obeying Maxwell's equations.  Hence, as usual, the two
longitudinal modes simply decouple, leaving the single scalar degree of
freedom.

The above analysis agrees completely with that of \cite{KLS}, which has
however been criticized as deviating from NGT by its use of second
order
form.  Let us therefore explicitly construct the canonical formulation
directly from the first order one, so that the degrees of
freedom may be counted directly.
Expanding (\ref{Ga}) to quadratic order in the fields we have
$(\sqrt{-g}g^{\mu\nu} \equiv \eta^{\mu\nu} + h^{\mu\nu} + B^{\mu\nu})$
\begin{equation}
{\cal L}^{(1)}_2 = (h^{\mu\nu} + B^{\mu\nu})
\partial_{[\lambda}\Gamma^\lambda_{\mu\nu ]} -
\eta^{\mu\nu}\:\Gamma^\alpha_{\mu [\lambda}\Gamma^\lambda_{\alpha\nu ]}
\label{Gj}
\end{equation}
where only $[\lambda\nu ]$ are antisymmetrized.  Decomposing
$\Gamma^\lambda_{\mu\nu}$
into its symmetric $(S)$ and antisymmetric $(A)$, parts, we obtain
\begin{eqnarray}
{\cal L}^{(1)}_2 & = & h^{\mu\nu} \:
\partial_{[ \lambda}S^\lambda_{\nu ] \mu} -
\eta^{\mu\nu} \: S^\alpha_{\mu [\lambda}S^\lambda_{\nu ] \alpha} - B^{\mu\nu}
 \:
\partial_{[\lambda}A^\lambda_{\nu ]\mu} + \eta^{\mu\nu}
A^\alpha_{\mu [\lambda}A^\lambda_{\nu ] \alpha}
\nonumber  \\
& - & h^{\mu\nu} \: \partial_\nu \:A^\lambda_{\mu\lambda} - B^{\mu\nu} \:
\partial_\nu \: S^\lambda_{\mu\lambda} + \eta^{\mu\nu}\:S^\alpha_{\mu\nu}
\:A^\lambda_{\alpha\lambda}
\;.
\label{Gk}
\end{eqnarray}
Only the last three terms in (\ref{Gk}) couple the symmetric and
antisymmetric
components, and do so only through traces of the affinity; hence we may
diagonalize just by shifting to $\widetilde{\Gamma}^\lambda_{\mu\nu} =
\Gamma^\lambda_{\mu\nu} + \frac{2}{D-1} \delta^\lambda_\mu \Gamma_\nu$, to
make $\widetilde{A}^\lambda_{\mu\lambda} = 0$.  Since $R_{\mu\nu}(\Gamma) =
R_{\mu\nu}(\widetilde{\Gamma}) - \frac{2}{D-1}\partial_{[\mu}\Gamma_{\nu ]}$,
where $\Gamma_\nu \equiv A^\lambda_{\nu\lambda}$, we get simply
${\cal L}^{(1)}_2  = {\cal L}^{(1)}_E + {\cal L}^{(1)}_B$ where
${\cal L}^{(1)}_E$ is linearized Einstein gravity in first order form
(known
of course to be equivalent to its second order one!) and
\begin{equation}
{\cal L}^{(1)}_B = B^{\mu\nu} \:
\partial_{\lambda}\widetilde{A}^\lambda_{\mu\nu} +
\eta^{\mu\nu} \: \widetilde{A}^\alpha_{\mu
\lambda}\widetilde{A}^\lambda_{\nu \alpha}
- \tilde{b}_\nu \partial_\mu B^{\mu\nu} \; ,
\label{Gl}
\end{equation}
in terms of $\tilde{b}_\mu \equiv -\frac{4}{D-1}\Gamma_\mu +
S^\lambda_{\mu\lambda}$.

The canonical analysis of (\ref{Gl}) is completely straightforward, and we
find
\begin{eqnarray}
{\cal L}^{(1)}_B & = & \pi^{ij} \: \partial_0 \: B_{ij} + \pi^i \:
\partial_0 B_{i0} - B^{ij} \: \partial_i \: \pi_j +
\tilde{b}_0\:\partial^i \: B_{i0} \nonumber \\
& + & \widetilde{A}^j_{0i} [2 \partial_j B_{0i} + 2\pi_{ij}] -
\widetilde{A}^j_{0i}\widetilde{A}^i_{0j} \nonumber \\
& + & \widetilde{A}^0_{0i} \left[ \left( -\textstyle{\frac{2}{D-2}}
-2\right)
\partial_j \: B^{ij} \right] +
\textstyle{\frac{D-1}{D-2}} (\widetilde{A}^0_{0i} )^{2} \nonumber \\
& + & \bar{A}^k_{ij} [-\partial_k B^{ij} ] - \bar{A}^k_{ij} \bar{A}^j_{ki} \; .
\label{Gm}
\end{eqnarray}
Here the conjugate momenta have been identified via
\begin{equation}
\pi_{ij} = - \widetilde{A}^0_{ij} \; , \;\;\; \pi_i = - \tilde{b}_i -
2\widetilde{A}^0_{0i} \; ,
\label{Gn}
\end{equation}
we have separated out the trace pieces
\begin{equation}
\widetilde{A}^k_{ij} = \bar{A}^k_{ij} +
\textstyle{\frac{1}{D-2}} \delta^k_{[j} A_{i]} \;
, \;\; A_i = \widetilde{A}^j_{ij} \; ,
\label{Go}
\end{equation}
and the $\widetilde{A}^\lambda_{\mu\lambda} = 0$ constraint has been applied.
The ``$p\dot{q}$" terms in (\ref{Gm}) show that
$(\widetilde{A}^j_{0i} \; , \; \widetilde{A}^0_{0i} \;, \; \bar{A}^k_{ij})$
are
all
auxiliary fields, while the canonical pairs $(\pi^{ij}, \; B_{ij})$,
$(\pi^i ,\;B_{0i})$ obey the one constraint $\partial^i B_{0i} = 0$,
enforced by the Lagrange multiplier $\tilde{b}_0$.  This means there are
$\frac{1}{2} (D-1)(D-2) + (D-1) - 1 = \frac{1}{2} (D+1)(D-2)$ degrees of
freedom, all completely equivalent to the second order analysis.

\end{document}